\documentclass[letterpaper,twocolumn,10pt]{article}
\usepackage{usenix-2020-09}

\usepackage{tikz}
\usepackage{amsmath}
\usepackage{booktabs}

\usepackage{graphicx}
\usepackage{caption}
\usepackage{subcaption} 
\graphicspath{{./figures/}}

\usepackage{filecontents}

\usepackage{tikz}
\usepackage{xcolor}

\usepackage{tabularx}
\usepackage{diagbox}
\usepackage{adjustbox}
\usepackage{hhline}
\usepackage{multirow}
\usepackage{makecell}

\usepackage{algorithm}
\usepackage{algorithmicx}
\usepackage{algpseudocode}

\usepackage{amsmath}
\usepackage[normalem]{ulem}
\usepackage[T1]{fontenc}

\begin{document}

\newcommand{\ada}[1]{\textcolor{red}{AG: #1}}
\newcommand{\ketan}[1]{\textcolor{blue}{KB: #1}}
\newcommand{\red}[1]{\textcolor{red}{#1}}
\newcommand{\green}[1]{\textcolor{green}{#1}}
\newcommand{\blue}[1]{\textcolor{blue}{#1}}
\newcommand{\jin}[1]{\textcolor{blue}{J: #1}}
\newcommand{\jmark}[1]{\textcolor{black}{#1}}

\newcommand{\tlide}{$\sim$}
\newcommand{\mul}{$\times$}
\newcommand{\eg}{\emph{e.g.}}
\newcommand{\ie}{\emph{i.e.}}
\newcommand{\etal}{\emph{et al.}}

\newcommand{\VHG}{81.71}
\newcommand{\VHM}{77.42}
\newcommand{\VHP}{44.84}

\newcommand{\HG}{65.42}
\newcommand{\HM}{59.64}
\newcommand{\HP}{42.57}

\newcommand{\MG}{51.01}
\newcommand{\MM}{47.31}
\newcommand{\MP}{38.1}

\newcommand{\LG}{24.43}
\newcommand{\LM}{22.76}
\newcommand{\LP}{16.98}

\newcommand{\summary}[1]{\noindent\textbf{#1}}

\newcommand*\circled[1]{\tikz[baseline=(char.base)]{
            \node[shape=circle,fill,inner sep=1pt] (char) {\textcolor{white}{#1}};}}

\newenvironment{tightitemize}%
 {\begin{list}{$\bullet$}{%
 		\setlength{\leftmargin}{10pt}
        \setlength{\itemsep}{0pt}%
        \setlength{\parsep}{0pt}%
        \setlength{\topsep}{0pt}%
        \setlength{\parskip}{0pt}%
        }%
 }%
{\end{list}}

\newcommand{\sys}{Stimpack}

\date{}

\title{\sys: An Adaptive Rendering Optimization System for Scalable Cloud Gaming}

\author{
{\rm Jin Heo$^\dagger$\thanks{Work done while the author was at Georgia Tech.}, Vic Wang$^\ddagger$, Ketan Bhardwaj$^\ddagger$, and Ada Gavrilovska$^\ddagger$}\\
\\
$^\dagger$Dolby Laboratories, $^\ddagger$Georgia Tech
}


\maketitle
\begin{abstract}
In distributed multimedia applications, content is often delivered to users in a degraded form due to network-induced lossy compression.
Real-time and interactive use cases like cloud gaming, which render content on the fly, require low latency and are hosted at resource-constrained edge servers.
We present a new insight: when rendered content is delivered over a network with lossy compression, high-quality rendering can be ineffective in improving user-perceived quality, leading to a poor return on computing resources.
Leveraging this observation, we built \sys, a novel system that adaptively optimizes game rendering quality by balancing server-side rendering costs against user-perceived quality.
The system uses a mechanism that quantifies the efficiency of resource usage to maximize overall system utility in multi-user scenarios.
Our open-sourced implementation and extensive evaluations show that \sys\ achieves up to 24\% higher service quality and serves twice as many users with the same resources compared to baselines.
A user study further validates that \sys\ provides a measurably better user experience.
\end{abstract}



\section{Introduction}
\label{sec:introduction}

Cloud gaming enables users to play video games on commodity devices by offloading game execution to a remote server and streaming the rendered output.
Unlike streaming pre-generated content, servers generate frames on the fly in response to real-time user interactions.
This requires a powerful server with graphics processing units (GPUs) and a low-latency network to deliver the high frames per second (FPS) and visual quality necessary for a responsive user experience.

While cloud gaming offers the benefit of playing video games without powerful user devices, its stringent network and computation requirements pose a fundamental challenge: the trade-off between latency and scalability.
To ensure a low-latency connection, game servers must be located close to users, typically at resource-constrained edge sites~\cite{meta:gameedge,deng2016server,shea2013cloud, heo2023flexr}.
However, this proximity creates a scalability bottleneck, as these servers are provisioned with limited resources and struggle to serve a number of concurrent users~\cite{soliman2013mobile,yami2020sara,chen2019t,grizan2015djay}.
While centralized datacenters can provide massive scalability, they are too far from most users to offer a low-latency connection, compromising service accessibility~\cite{choy2012brewing,sabet2020latency,graff2021analysis}.

To ensure desired service quality, existing cloud gaming services limit session playtime and queue users when nearby servers are fully occupied~\cite{geforcenow,xboxcloudgaming}.
While this approach guarantees service quality, it significantly diminishes service availability with long wait times.
Previous research has explored serving multiple users by focusing on network resources, such as through effective bandwidth allocation and adaptive compression~\cite{chen2019t,wang2013adaptive,slivar2019qoe,hong2015enabling,yami2020sara}.
However, they do not offer a fundamental solution for serving concurrent users on edge servers with limited computational resources.
Although a separate line of research has explored reducing the server's computational costs and maintaining playable FPS by adjusting the game rendering quality (RQ)~\cite{grizan2015djay}, it does not consider how the generated content's quality is actually realized on the user side, given the effects of network-induced lossy compression.
This can lead to suboptimal user experiences and inefficient resource usage.

\begin{figure}[!h]
  \centering
  \vspace{-2.0ex}
  \includegraphics[width=0.87\linewidth]{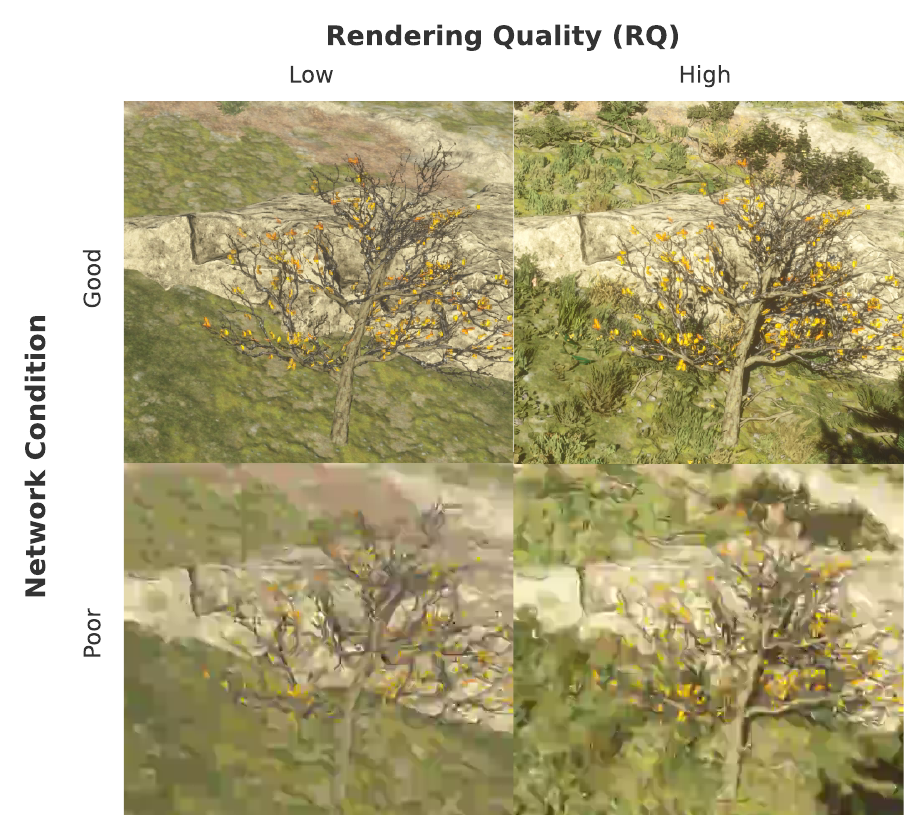}
  \vspace{-1.0ex}
  \caption{The screenshots of a sample game scene with different RQs and compression parameters of H.264 corresponding to good and poor network conditions}
  \label{fig:motiv}
  \vspace{-2.0ex}
\end{figure}

We identify a new insight: the impact of a server's computational effort on user-perceived quality varies significantly depending on the user's network condition and the resulting degree of lossy compression.
When a user's network is poor, frames generated with the highest RQ can fail to improve the user's perceived quality due to heavy compression.
Concretely, in Figure~\ref{fig:motiv}, the highest RQ takes \tlide5.4 ms to render the sample scene, while the lowest takes just \tlide1.5 ms on our testbed (\S\ref{sec:expsetup}).
In cases of severe compression loss (bottom row of Figure~\ref{fig:motiv}), the highest RQ, which takes 3.6\mul\ more rendering time, becomes ineffective as its quality improvement is overshadowed by compression loss.
In contrast, with a good network connection (top row), the resource usage for the higher RQ is better justified as it is effective in improving the user-side visual quality.

This observation provides a fundamental opportunity to systematically utilize server resources more efficiently and maintain user experiences.
Our goal is to mitigate the scalability issue on resource-constrained servers at the edge of the network by accommodating more users with each GPU.
To realize this, we introduce \textbf{\sys}, a novel system that adaptively optimizes the resources used for rendering game content.
\sys\ operates by quantifying resource efficiency based on the rendering cost (\ie, rendering time) and the user-side visual quality, which is estimated by considering both RQ and compression lossiness.
By systematically leveraging this metric, \sys\ adapts each user's RQ to ensure that resource usage is effectively translated into user-perceived quality while maintaining a playable FPS, thereby maximizing overall system utility in multi-user scenarios.

\sys's adaptive RQ optimization is enabled by three key features.
(1) It estimates user-side visual quality on the server by using a prediction model that considers both a given RQ and compression parameter (\S\ref{sec:frameprediction}).
(2) \sys\ introduces a scoring mechanism that quantifies the efficiency of an RQ setting with respect to its rendering cost and estimated visual quality.
This allows \sys's round-based RQ optimization process to systematically balance the rendering cost required for a playable FPS against the user's perceived quality.
Additionally, in multi-user scenarios, it helps to prioritize and coordinate RQ adjustments among users to maximize overall system utility (\S\ref{sec:rq_opt}).
(3) As changing RQ incurs overheads due to the reconfiguration and reloading of graphical assets on the GPU, \sys\ adopts a backoff mechanism.
This mitigates the negative impact of frequent and oscillatory RQ adjustments on user experience and helps to stabilize each user's RQ at a suitable level (\S\ref{sec:rq_oscillation}).

Overall, this paper makes the following contributions:
\begin{tightitemize}
  \item We present a new insight that a server's computational effort can be wasted due to network-induced lossy compression, and demonstrate that user-perceived quality can be estimated from server-side information.
  \item We introduce \sys, a novel system that adaptively optimizes multi-user RQ settings based on its efficiency score and incorporates a stabilization mechanism to ensure a stable user experience.
  \item We validate \sys's effectiveness through comprehensive evaluations with two Unreal Engine-based games and a user study.
  \item We open-source \sys\footnote{\url{https://github.com/gt-stimpack/Stimpack}}, hoping to reduce the barriers for further research.
\end{tightitemize}

\section{Background}
\label{sec:background}

\begin{figure}[t]
  \centering
  \includegraphics[width=0.87\linewidth]{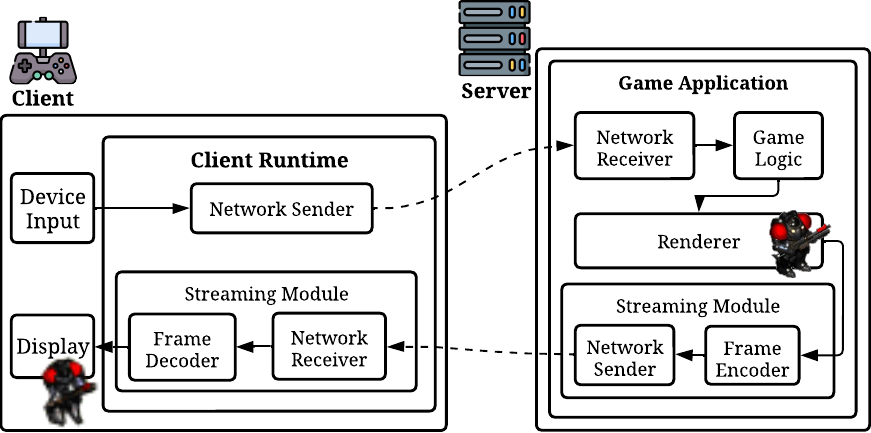}
  \vspace{-0.5ex}
  \caption{General architecture of cloud gaming}
  \label{fig:cloud_gaming}
\end{figure}

\noindent\textbf{Cloud Gaming.}\quad
Cloud gaming services use a client-server architecture to offload game execution to a remote server.
As shown in Figure~\ref{fig:cloud_gaming}, the client runtime on the user side captures user inputs and sends them to the server.
The game application, running on the server, processes these inputs and renders the game contents.
For efficient transmission of the generated game frames, the streaming module on the server uses a video codec, \eg, H.264~\cite{wiegand2003overview} and HEVC~\cite{sullivan2012overview}, which performs lossy compression.
The encoded frames are transmitted to the client via streaming methods, \eg, WebRTC~\cite{webRTC} and RTP with RTCP~\cite{schulzrinne2003rtp}.
Streaming modules estimate the available bandwidth using congestion control algorithms such as Google congestion control~\cite{holmer2015google} and adapt its compression parameter to the estimated bandwidth~\cite{chen2019t}.

\begin{table}[]
  \caption{\label{tab:rqqp_summary} Summary of rendering quality (RQ) and network conditions' compression parameter (QP)
  settings in this paper}
  \vspace{-1.0ex}
  \centering
  \resizebox{0.95\linewidth}{!}{
    \begin{tabular}{|c|c|}
    \hline
      Term           & Description   \\ \hline
      RQ             & \makecell{ The quality levels with different intensity of rendering optimizations \\ \texttt{Low}, \texttt{Medium}, \texttt{High}, and \texttt{Very High} (the higher, the better quality)} \\ \hline
      QP             & \makecell{ The lossiness parameter to control the compression ratio \\ \texttt{10} (Good), \texttt{30} (Fair), \texttt{40} (Poor) (the higher, the more lossy)} \\ \hline
\end{tabular}
  }
  \vspace{-1.5ex}
\end{table}

\noindent\textbf{Rendering Optimization and Quality.}\quad
3D graphics are generated through a rendering pipeline, a multi-stage process that includes vertex processing, geometry, and pixel shaders to render 3D objects and convert them into a 2D image.
Throughout the pipeline, there are optimization opportunities, \eg, visibility and distance-based culling, anti-aliasing, and texture mipmaps.
These optimizations introduce a trade-off between generated content quality and computational cost.
Modern game engines, \eg, Unity~\cite{unity}, Unreal Engine~\cite{unreal}, and Godot~\cite{godot}, provide a knob to adjust the RQ during gameplay.
These engines offer preset options that curate optimization levels across the entire rendering pipeline; we have further discussion on the RQ control granularity in Appendix~\ref{sec:rq_granularity}.

\begin{figure}[t]
  \centering
  \begin{subfigure}[t]{0.45\textwidth}
    \includegraphics[width=\textwidth]{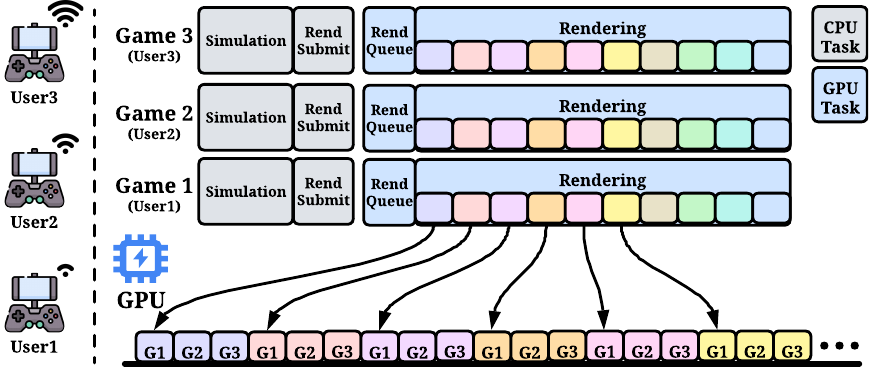}
    \caption{Multiple game application rendering on a GPU}
    \label{fig:vis1}
  \end{subfigure}\hspace{0.2\textwidth}
  \begin{subfigure}[t]{0.45\textwidth}
    \includegraphics[width=\textwidth]{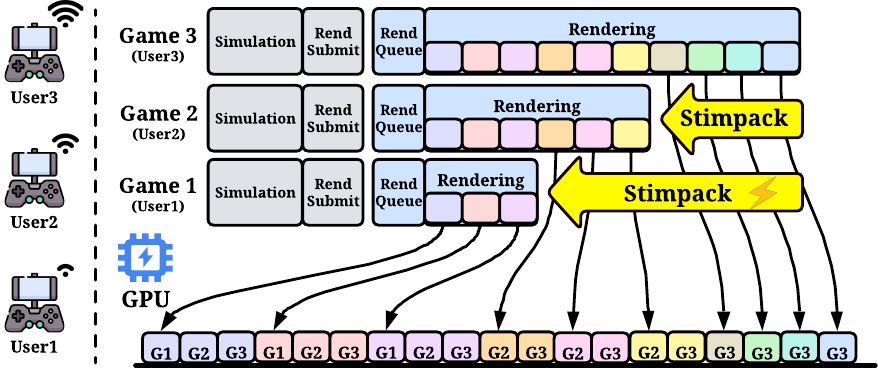}
    \caption{\sys\ adapts the rendering workloads via RQ optimization with the consideration on resource efficiency and user-side service quality.}
    \label{fig:vis2}
  \end{subfigure}
  \vspace{-1.0ex}
  \caption{The visualization of multi-application rendering on a GPU}
  \label{fig:vis}
  \vspace{-3.0ex}
\end{figure}

\noindent\textbf{Rendering Multiple Game Applications on GPU.}\quad
Modern GPUs and drivers use time-multiplexing to share GPU resources among multiple applications.
Figure~\ref{fig:vis1} shows how multiple game applications can be rendered on a single GPU.
The CPU runs the game simulation and submits rendering commands to the GPU.
The GPU then processes these commands in a time-multiplexed manner, switching between the rendering queues of different applications.
Therefore, in cloud gaming, serving more users on a shared GPU directly increases rendering latency, which in turn degrades the user experience with lower FPS and reduced responsiveness.

\noindent\textbf{Rendering Quality and Compression Parameter Settings.}\quad
In this paper, we use the optimization presets of Unreal Engine to define four levels of RQ: \texttt{Low}, \texttt{Medium}, \texttt{High}, and \texttt{Very High}, as summarized in Table~\ref{tab:rqqp_summary}.
For each RQ level, rendering optimization techniques are applied with different degrees of intensity.
For example, when the RQ is Low, optimizations are applied most aggressively to achieve the lowest visual quality and computational cost.

Similarly, video codecs determine compression rates using a lossiness parameter; H.264 and HEVC employ a quantization parameter (QP) that ranges from 0 (lossless) to 51 (most lossy).
We establish three QP settings -- 10 for good network conditions, 30 for fair, and 40 for poor -- corresponding to different degrees of compression, as also shown in Table~\ref{tab:rqqp_summary}.

\section{Motivation}
\label{sec:motivation}
In cloud gaming, the user-perceived visual quality is influenced by both RQ and network-dictated QP.
Lowering the RQ reduces the rendering workload for each user, which allows a shared GPU to accommodate more users while maintaining playable FPS, as shown in Figure~\ref{fig:vis}.
However, blindly setting the lowest RQ to reduce cost can significantly compromise user experiences.
Therefore, it is crucial to determine the effective RQ for each user, striking a balance between computational efficiency and maintaining a user experience.

\begin{figure}[t]
  \centering
  \includegraphics[width=0.7\linewidth]{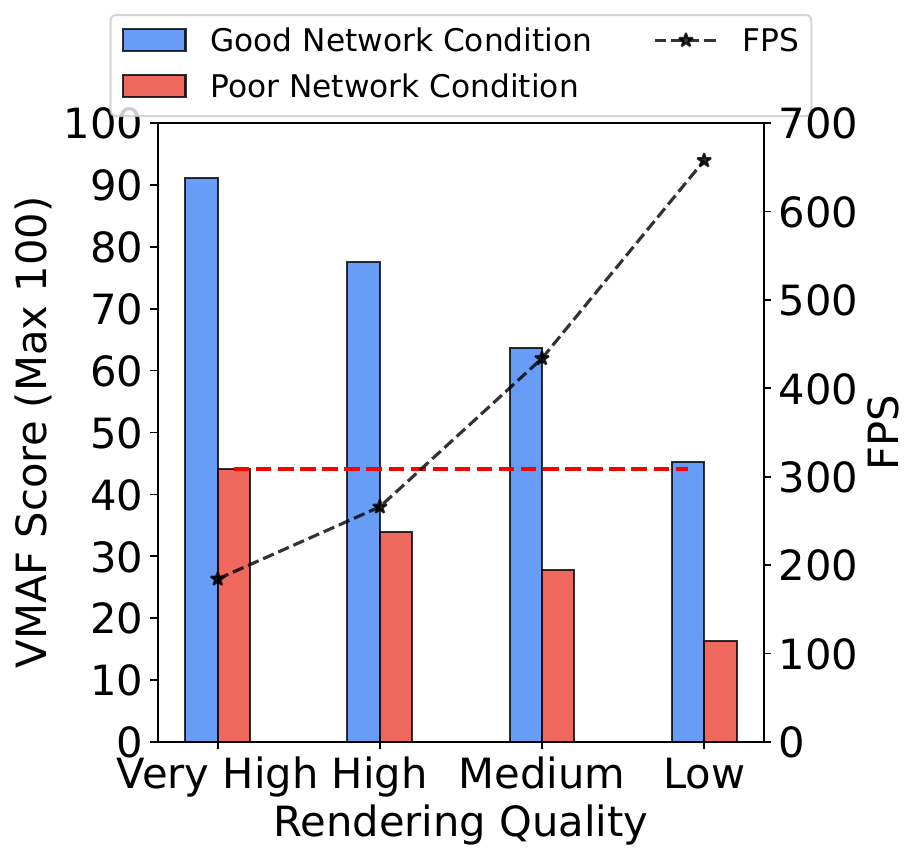}
  \vspace{-1.0ex}
  \caption{The FPS and visual quality (VMAF) measurements of the scene in Figure~\ref{fig:motiv} with different RQs and QPs}
  \label{fig:motiv_rq_cost}
  \vspace{-2.0ex}
\end{figure}

Our key observation is that the quality loss due to compression is more significant for frames of higher RQ than those of lower RQ, resulting in lower efficiency.
As introduced, Figure~\ref{fig:motiv} shows the qualitative results of this visual quality loss.
For a given scene, when a network condition is poor, the details of a higher-RQ frame become blurred and difficult to appreciate due to severe compression loss, even though more rendering resources were used.

This insight is further supported by quantitative results from our testbed, as shown in Figure~\ref{fig:motiv_rq_cost}.
A sample scene shows that the highest RQ takes 3.6\mul\ more per-frame rendering time than the lowest RQ (5.4 ms vs. 1.5 ms).
Visual quality, measured by VMAF~\cite{vmaf} ranging from 0 (bad) to 100 (excellent), degrades significantly more for \texttt{Very High} RQ due to network conditions: from 91.1 (good) to 45.3 (poor).
In contrast, \texttt{Low} RQ shows less quality loss, degrading from 45.1 (good) to 16.4 (poor).
Based on the interpretation guide of VMAF score~\cite{vmaf}, the quality loss is more severe for \texttt{Very High} RQ.
Importantly, as highlighted with the red dotted line, \texttt{Very High} RQ under poor network conditions yields a similar quality (45.3) to \texttt{Low} RQ under good network conditions (45.1), while still taking 3.6\mul\ more per-frame rendering time.
These findings highlight the need to strategically assign RQ for each user by balancing rendering costs with the user-perceived visual quality of RQ under network conditions.

\section{\sys}
\label{sec:stimpack}

Based on the insights from our motivation, we introduce \sys, a novel system designed to adaptively optimize RQ for scalable cloud gaming.
\sys\ aims to improve a game server's scalability by optimizing rendering workloads to balance resource efficiency with user experiences.
To achieve this, \sys\ develops an efficiency score by leveraging a prediction of user-perceived visual quality (based on RQ and QP) and monitored per-frame rendering latency.
This score guides its multi-user RQ optimization process.
Furthermore, \sys\ incorporates a stabilization mechanism to mitigate the negative impacts of frequent and oscillatory RQ adjustments on user experiences.

\subsection{System Overview}
\label{sec:overview}


\begin{figure}[t!]
  \includegraphics[width=\linewidth]{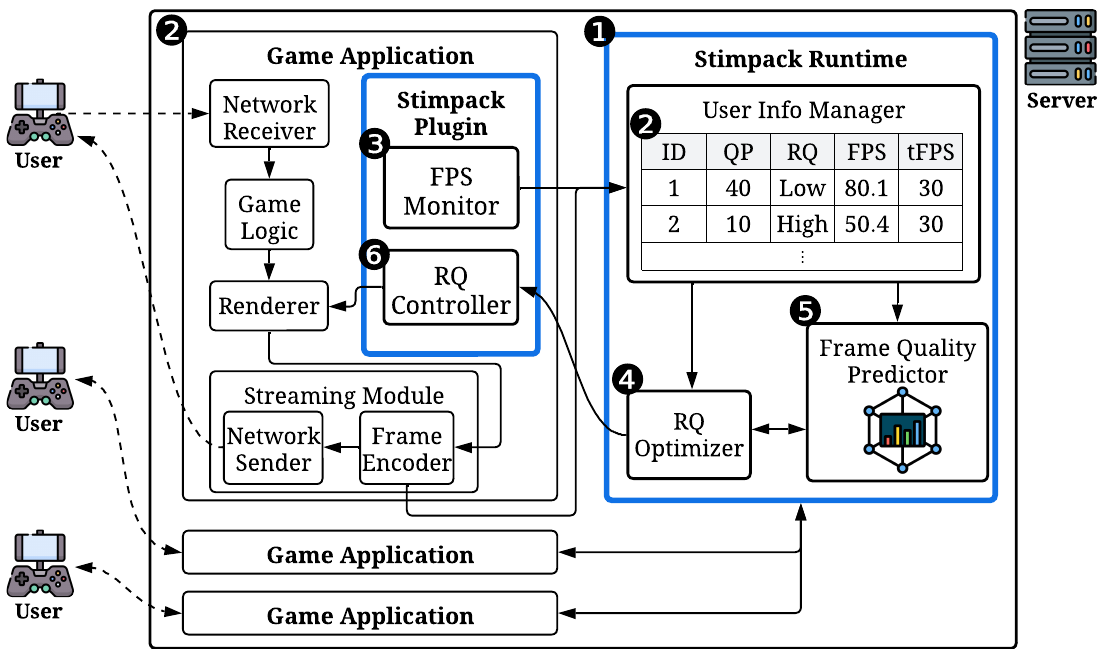}
  \caption{The architecture of \sys}
  \label{fig:sys}
  \vspace{-2.0ex}
\end{figure}

As shown in Figure~\ref{fig:sys}, \sys\ is a server-side system with two main components: the \emph{\sys\ plugin} and the \emph{runtime}.
The plugin integrates into each game application, while the runtime system orchestrates the overall optimization process.

\circled{1} The server-side runtime includes a user information manager and an RQ optimizer with a quality predictor for estimating user-perceived visual quality.
\circled{2} When a game starts, its \sys\ plugin initializes user information, including user ID and threshold FPS (tFPS) for playable experience, to the user information manager.
QP is fetched from the streaming module; \sys\ uses compression parameters, making it agnostic to streaming methods.
It sets the initial RQ to the lowest level to reduce initial lag and ensure higher FPS than the threshold, also setting an FPS upper bound to prevent FPS from going too high to overuse server resources.

\circled{3} The \sys\ plugin periodically monitors per-frame rendering latencies and updates the current FPS to the runtime.
\circled{4} The RQ optimizer is round-based and adjusts RQ based on the updated information.
In each adjustment round, the optimizer promotes or demotes the RQ of a user with the highest priority to maximize the overall efficiency and maintain playable status.
\circled{5} When the RQ optimizer finds the highest-priority user, the priority is determined by the efficiency score (\S\ref{sec:scoring}).
The efficiency score is calculated as a weighted sum of the per-frame rendering latency and the predicted user-side visual quality.
It uses a pretrained regression model to predict the user-side visual quality with the given RQ and QP (\S\ref{sec:frameprediction}).

\circled{6} When the optimizer makes a decision for an RQ change, it is forwarded to the plugin of the corresponding game application.
The RQ controller then changes its RQ.

\begin{figure*}[ht!]
  \centering
  \begin{subfigure}[t]{0.18\textwidth}
    \includegraphics[width=\textwidth]{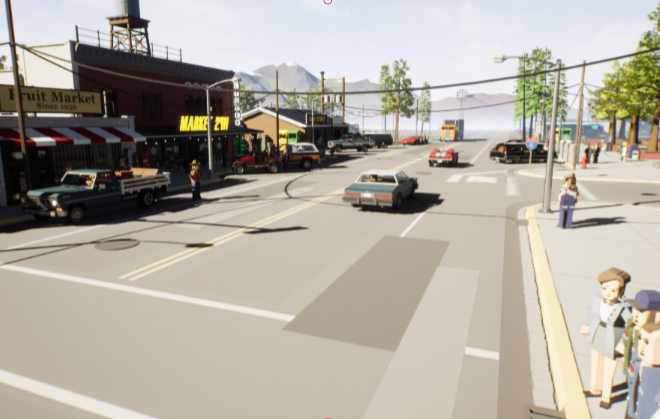}
    \vspace{-3.0ex}
    \caption{Town}
    \label{fig:sc1}
  \end{subfigure}\hspace{0.01\textwidth}
  \begin{subfigure}[t]{0.18\textwidth}
    \includegraphics[width=\textwidth]{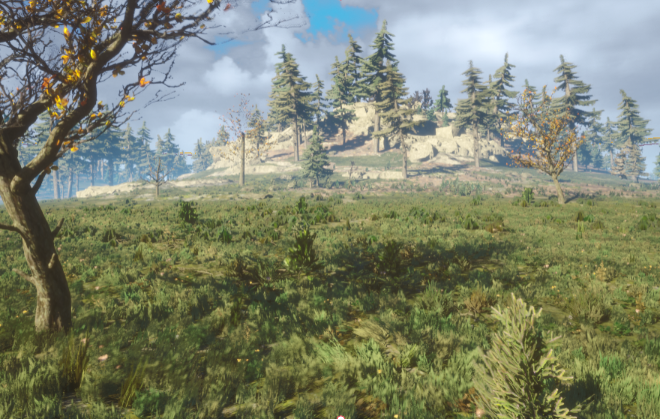}
    \vspace{-3.0ex}
    \caption{Forest}
    \label{fig:sc2}
  \end{subfigure}\hspace{0.01\textwidth}
  \begin{subfigure}[t]{0.18\textwidth}
    \includegraphics[width=\textwidth]{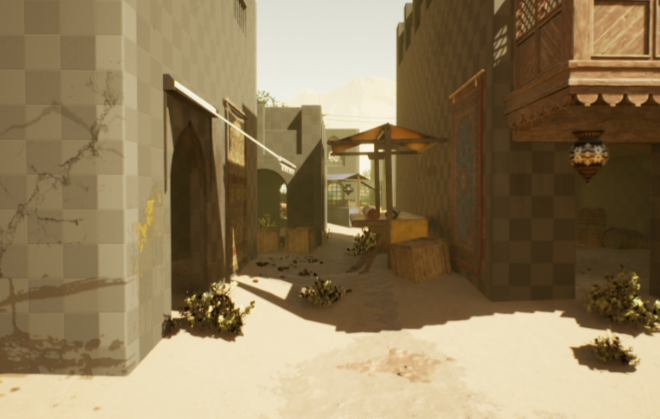}
    \vspace{-3.0ex}
    \caption{Desert}
    \label{fig:sc3}
  \end{subfigure}\hspace{0.01\textwidth}
  \begin{subfigure}[t]{0.18\textwidth}
    \includegraphics[width=\textwidth]{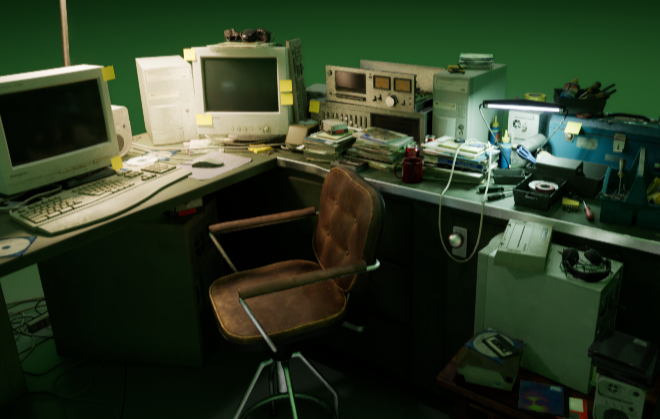}
    \vspace{-3.0ex}
    \caption{Office}
    \label{fig:sc4}
  \end{subfigure}\hspace{0.01\textwidth}
  \begin{subfigure}[t]{0.18\textwidth}
    \includegraphics[width=\textwidth]{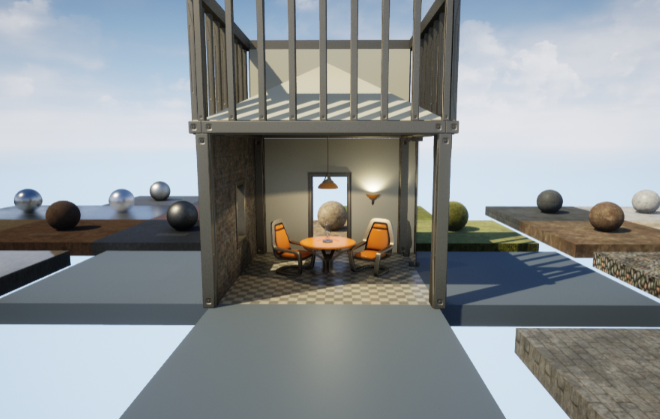}
    \vspace{-3.0ex}
    \caption{Sky Field}
    \label{fig:sc5}
  \end{subfigure}
  \vspace{-2.0ex}
  \caption{The screenshots of 3D scenes used for predicting the frame quality with given RQs and QPs}
  \label{fig:sc}
  \vspace{-2.0ex}
\end{figure*}

\subsection{User-side Visual Quality Prediction}
\label{sec:frameprediction}
\sys's design is predicated on the premise that a server can accurately estimate a user-perceived visual quality.
To validate this, we need to address two key questions: how to design a prediction model, and whether its performance is sufficient.
We approach this prediction task as a regression problem, where the model predicts the visual quality metric, VMAF, based on server-side parameters, \ie, RQ and QP.
Through training and testing regression models, we demonstrate the feasibility and accuracy of estimating user-side visual quality from the server.

\vspace{-1.0ex}
\begin{equation}\label{eq:vq}
  \small
  \textit{VMAF}_{est} = f\left(RQ_u, QP_u\right)
\end{equation}
\vspace{-1.0ex}

\summary{Training and Testing Data.}\quad
The regression problem is formulated in Eq.~\ref{eq:vq}, where \(f\) is the regression model, and an estimated VMAF is in the range of \(\left[0, 100\right]\).
To train and test \(f\), we collected 60,000 frames recorded from different 3D scenes with the combination of RQ and QP settings.
The scenes are Town, Forest, Desert, Office, and Sky Field in Figure~\ref{fig:sc}.
In each scene, we set 5 different locations without overlapping field of views to prevent the test data from being leaked to the training data.
From each location, we collected 150 frames for each RQ: \texttt{Low}, \texttt{Medium}, \texttt{High}, and \texttt{Very High}.
The recorded frames are encoded and decoded with QP settings for good, fair, and poor network conditions described in Table~\ref{tab:rqqp_summary}.

As the ideal case is with the original frames of \texttt{Very High} RQ without compression loss, we use them as the reference for VMAF calculation.
The VMAF data is generated by comparing the frames of different RQs and QPs to the reference frames.
The training and test data are split based on the capturing locations in the scenes; among the 5 locations, 4 are for training data and the remaining location is for testing data.

\summary{Prediction Models and Errors.}\quad
Using our synthesized data, we trained and tested several regression models to predict \(\textit{VMAF}_{est}\) based on RQ and QP.
To select the most effective model, we benchmarked different linear and non-linear regressors, including support vector (SVR), linear, K-nearest neighbor (KNN), decision tree, AdaBoost, and Bagging.
As shown in Table~\ref{tab:regression}, we calculated the prediction errors as the root-mean-square error (RMSE) between predicted and ground-truth values.
The decision tree regressor achieved the lowest error with an RMSE of 9.05, a reasonably low value considering the VMAF scale of [0-100].
In addition to these results, we conducted cross-validation with different train-test splits, confirming that user-perceived visual quality can be estimated with reasonable accuracy (Appendix~\ref{sec:cross_validation}).

\begin{table}[t]
  \caption{\label{tab:regression} RMSE of different regressors for VMAF prediction}
  \vspace{-1.0ex}
  \centering
  \resizebox{\linewidth}{!}{
    \begin{tabular}{|l|c|c|c|c|c|c|} \hline
                        & SVR     &  Linear & KNN             & DecisionTree   & AdaBoost    & Bagging       \\ \hline
    RMSE                & 18.36   &  11.88  & 10.7            & \textbf{9.05}  & 12.26       & 9.05         \\ \hline
\end{tabular}
  }
  \vspace{-2.0ex}
\end{table}

\subsection{Rendering Quality Optimization}
\label{sec:rq_opt}
\summary{Objectives and Approach.}\quad
\sys's primary goal is to improve resource efficiency by optimizing RQ, thereby enhancing a game server's scalability and accommodating more users while maintaining their gaming experience.
Achieving this requires an adaptive RQ optimization with three key objectives.
\textbf{(O1)} The system must maintain users' FPS above a threshold (tFPS) to ensure a minimum bar for a playable experience.
\textbf{(O2)} With this constraint met, the system should maintain user-perceived visual quality to deliver a satisfactory and immersive experience.
\textbf{(O3)} In multi-user scenarios, the optimization process needs to prioritize and coordinate RQ adjustments across users to maximize system-wide utility.
To accomplish these objectives, \sys\ employs a round-based RQ optimization process based on its scoring mechanism that quantifies server resource efficiency.

\subsubsection{Efficiency Score for Multi-user Optimization}
\label{sec:scoring}
The efficiency score is a key component that enables the prioritization and coordination of RQ optimization in multi-user scenarios.
Each user's efficiency score, \(Score_u\) (Eq.~\ref{eq:score}), is defined as a unified metric that quantifies server resource efficiency by weighing the rendering cost, which is based on the monitored FPS, against the predicted user-perceived visual quality, which is based on the estimated VMAF.
Both score terms, \(\textit{FPS}\_Score_u\) (Eq.~\ref{eq:fps_score}) and \(\textit{VQ}\_Score_u\) (Eq.~\ref{eq:vq_score}), have a value between 0 and 1.
With this metric, \sys\ can systematically determine priorities and coordinate RQ adjustments among multiple users to maximize overall system efficiency.


\vspace{-1.0ex}
\begin{equation}\label{eq:score}
  \small
  \begin{aligned}
    &Score_u = \alpha\left(\textit{VQ}\_Score_u\right) + (1-\alpha)\left(\textit{FPS}\_Score_u\right), \\
    &\text{where}\quad 0 \leq \alpha \leq 1
  \end{aligned}
\end{equation}
\vspace{-1.0ex}

\(\textit{FPS}\_Score_u\) (Eq.~\ref{eq:fps_score}) is logarithmically scaled based on the current FPS and upper bound.
As FPS increases closer to the upper bound (lower cost), the score approaches 1.
The motivation of the logarithmic scaling is the human perception characteristics~\cite{reichl2010logarithmic, ou2010perceptual, ramamurthy2015human}.
For instance, an FPS increment is more effective at lower FPS ranges, with its impact diminishing as FPS increases;
the same 30 increment from 10 to 40 FPS is more noticeable than from 90 to 120 FPS.

\vspace{-0.5ex}
\begin{equation}\label{eq:fps_score}
  \small
    \textit{FPS}\_Score_u = max\left(0, 1 + log\left(\frac{\textit{FPS}_u}{\textit{FPS}_{upper}}\right)\right)
\end{equation}
\vspace{-0.5ex}

The visual quality score, \(\textit{VQ}\_Score_u\) (Eq.~\ref{eq:vq_score}), is a normalized value representing \(\Delta \textit{VMAF}_{est}\), the expected change in user-perceived quality due to an RQ adjustment.
This change quantifies the predicted quality gain from an RQ promotion or the quality loss from an RQ demotion.
We normalize this change by scaling \(\Delta \textit{VMAF}_{est}\) by the total predicted VMAF range (the difference between \(\textit{VMAF}_{est\_max}\) and \(\textit{VMAF}_{est\_min}\)).
This approach ensures that the visual quality score is also bounded between 0 and 1, making it a comparable term in the overall efficiency score.


\vspace{-1.5ex}
\begin{equation}\label{eq:vq_score}
  \small
  \begin{aligned}
    &\textit{VQ}\_Score_u = \frac{\Delta \textit{VMAF}_{est}}{\textit{VMAF}_{est\_max} - \textit{VMAF}_{est\_min}}, \\
    &\text{where}\quad \Delta \textit{VMAF}_{est} = \lvert f\left(RQ_{to\_change}, QP_u\right)-\textit{VMAF}_{est}\rvert
  \end{aligned}
\end{equation}
\vspace{-1.5ex}

\subsubsection{RQ Optimization Process}
\sys's RQ optimization process is designed to achieve the key objectives outlined earlier while accommodating multiple users given the available resources.
With \(Score_u\), the objective function for the optimization can be formulated as Eq.~\ref{eq:obj_func}.
The optimization process adapts users' RQ within the RQ levels described in Table~\ref{tab:rqqp_summary} to maximize the aggregate efficiency score across all users (\textbf{O2}),
while ensuring that each user's FPS remains within a playable range between the threshold and an upper bound (\textbf{O1})

The optimization in \sys\ is a round-based process that is performed periodically.
In each round, it adjusts the most-prioritized user's RQ by one level, either promoting or demoting, based on the current serving status.
This gradual adjustment helps prevent sudden and negative impacts on user experiences.
When \sys\ promotes a user's RQ, it selects the user with the highest \(Score_u\) because this indicates low rendering cost (high FPS) and an estimated high visual quality gain.
Conversely, for a demotion, \sys\ selects the user with the lowest \(Score_u\) because this means a high rendering cost and a low quality loss from lowering the RQ.
By adjusting the highest-priority user and then re-evaluating the serving status in the next round, \sys\ effectively coordinates RQ changes in a multi-user environment (\textbf{O3}).

\vspace{-1.5ex}
\begin{equation}\label{eq:obj_func}
  \small
  \begin{aligned} 
           &\max_{RQ} \sum_{u} Score_u,\\
           &\text{s.t.}\ \textit{FPS}_{thresh} \leq \textit{FPS}_u \leq \textit{FPS}_{upper}\quad\forall u \\
           &\quad\ RQ_{min} \leq RQ_u \leq RQ_{max}\quad\forall u
  \end{aligned}
\end{equation}
\vspace{-1.5ex}

Algorithm~\ref{alg:cap} describes \sys's RQ optimization process, which is invoked by the RQ optimizer.
The process first identifies candidates for demotion and promotion based on users' current RQs and the defined RQ levels.
The system prioritizes the demotion phase to maintain all users' FPS above the threshold.
This is because ensuring a playable FPS is a critical requirement for a satisfactory gaming experience; there is a physiological threshold for human perception to recognize motion portrayal~\cite{adobe24}, and game responsiveness can be significantly impaired with infrequent frame updates.
When there are users with FPS below the threshold, it selects the user with the lowest \(Score_u\) for demotion.

Once all users are served with an FPS above the threshold, the process enters the promotion phase to improve visual quality.
In this phase, it firstly checks if there are users with FPS close to the threshold among the promotion candidates.
If such users exist, the promotion is skipped to prevent the FPS drop below the threshold due to the promotion.
Otherwise, it calculates \(Score_u\) for promotion candidates and selects the user with the highest score.

\begin{algorithm}[t]
  \caption{The RQ optimization process of \sys}\label{alg:cap}
  \scriptsize
  \begin{algorithmic}[1]
    \Procedure{\text{Optimize\_RQ}}{user\_table, N = current round}
    \If {N < Backoff Round}
      \State \textbf{return}
    \EndIf

    \State \text{demote\_candidates} $\gets$ \text{users (RQ$_{min}$ $<$ RQ$_{\textit{u}}$ $\leq$ RQ$_{max}$)}
    \State \text{promote\_candidates} $\gets$ \text{users (RQ$_{min}$ $\leq$ RQ$_{\textit{u}}$ $<$ RQ$_{max}$)}

    \If {users (FPS$_{\textit{u}}$ < FPS$_{\textit{thresh}}$) exist \textbf{in} user\_table}
      \State{\bf/* RQ Demotion */}
      \For {candidate \textbf{in} demote\_candidates}
        \State {Calculate\ \textit{Score$_{u}$} (Eq.~\ref{eq:score})}
      \EndFor
      \State {demote\_user $\gets$ candidate\ of minimum \textit{Score$_{u}$}}
      \State {Demote\_RQ(demote\_user)}
      \State {adjusted\_user $\gets$ demote\_user}

    \Else
        \State{\bf/* RQ Promotion */}
        \If {candidates (FPS$_{\textit{u}}$ $<$ FPS$_{\textit{thresh}}$ + FPS$_{\textit{buffer}}$) exist}
          \State \textbf{return}
        \EndIf
        \For {candidate \textbf{in} promote\_candidates}
          \State {Calculate\ \textit{Score$_{u}$} (Eq.~\ref{eq:score})}
        \EndFor
        \State {promote\_user $\gets$ candidate\ of maximum \textit{Score$_{u}$}}
        \State {Promote\_RQ(promote\_user)}
        \State {adjusted\_user $\gets$ promote\_user}
    \EndIf

    \If {Is\_RQ\_oscillating(adjusted\_user)}
      \State Backoff Round $\gets$ {Get\_backoff\_round(N)}
    \EndIf
    \EndProcedure
  \end{algorithmic}
\end{algorithm}

\subsection{RQ Stabilization Mechanism}
\label{sec:rq_oscillation}
\begin{figure}[t!]
  \centering
  \includegraphics[width=\linewidth]{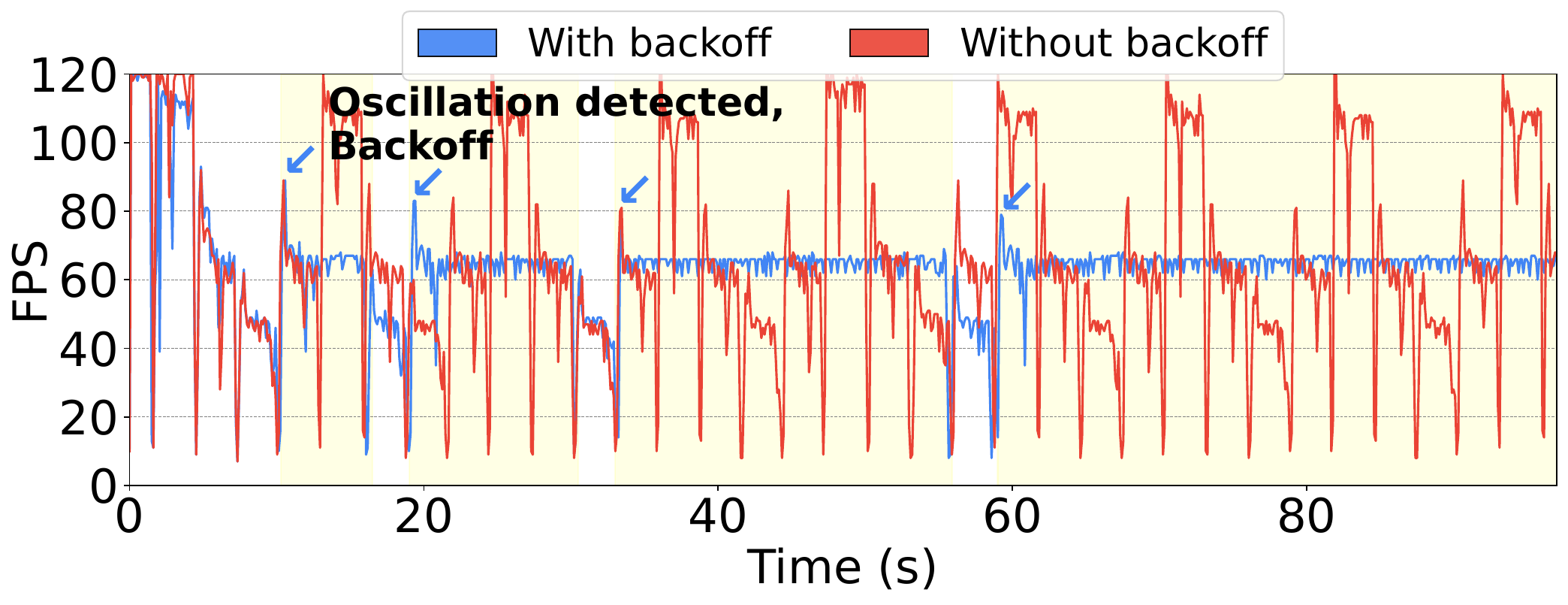}
  \vspace{-4.0ex}
  \caption{The FPS traces with and without the backoff}
  \label{fig:backoff}
  \vspace{-2.0ex}
\end{figure}

\summary{RQ Oscillation Issue.}\quad
As the optimization process adjusts a user's RQ by one level each round, it can cause the user's RQ to oscillate between two levels, leading to unnecessary RQ adjustments.
Such frequent and oscillatory RQ changes can significantly degrade the user's gaming experience.
This is because changing RQ presents overhead, as the rendering assets on a GPU need to be reloaded and reconfigured for the new RQ.
These FPS drops (downward spikes in Figure~\ref{fig:backoff}) are a direct result of this overhead.
When a user's RQ starts to oscillate and the optimization process is frequently invoked, the user experiences FPS drops and game lag every round, resulting in a poor gaming experience.

\summary{RQ Stabilization Mechanism with Backoff.}\quad
A naive solution to this problem is to set a long interval, but this prevents the RQ optimization process from adapting quickly to user state changes.
To address this, \sys\ has a stabilization mechanism based on exponential backoff.
This mechanism keeps track of a user's past RQ updates and detects when their RQ starts to oscillate.
If oscillation is detected, it sets a backoff round for the optimization process, which is then skipped until the current round reaches the backoff round.

The backoff round is set by the \textit{Get\_backoff\_round} function in Algorithm~\ref{alg:cap}, which increases the user's backoff count and sets the backoff round to \(N + \textit{backoff\_base}^{\textit{backoff\_count}}\); \textit{N} is the current round number.
\textit{backoff\_base} and \textit{backoff\_count} are adjustable parameters.
Additionally, to prevent excessively long suppression, \sys\ allows setting a maximum backoff count.
When a user's RQ stops oscillating, the backoff count is reset, and the optimization process is resumed.

To demonstrate the effectiveness of our stabilization mechanism, we conducted an experiment with an optimization interval of 3 seconds, a backoff base of 2, a maximum backoff count of 5, and an FPS threshold of 60.
Figure~\ref{fig:backoff} shows the FPS traces of users with and without the backoff.
The trace without the backoff shows that the RQ oscillation occurs frequently, causing frequent FPS drops.
Moreover, when the RQ change overhead affects the next round with low FPS, the user’s RQ is demoted again, leading to another FPS drop and poor visual quality.
Conversely, the user with the backoff initially experiences RQ oscillation, but the backoff relieves it after a few rounds, stabilizing the user’s RQ and providing a consistent gaming experience.
Our stabilization mechanism effectively relieves performance degradation caused by RQ oscillations, ensuring a stable gaming experience.

\section{Implementation}
\label{sec:implementation}
\sys, currently implemented and tested on Ubuntu 22.04, uses Python for its runtime and scikit-learn~\cite{pedregosa2011scikit} for regression models.
While it primarily supports Unreal Engine (UE) games through its plugin for UE, it can be easily adapted for other game engines like Unity~\cite{unity} and Godot~\cite{godot}.
The evaluation games are C++ implementations on UE, and ZeroMQ~\cite{zeromq} is used inter-process communication between the \sys\ runtime and plugin.

\section{Evaluation}
\label{sec:evaluation}
This section presents a comprehensive evaluation to demonstrate the feasibility and effectiveness of \sys.
Our evaluation addresses two key questions.
First, can \sys\ effectively increase a server's scalability by accommodating more users while maintaining playable FPS?
Second, compared to other baselines, does \sys\ better maintain gaming service quality and user experience beyond simply serving more users?
We answer these questions by conducting a series of experiments on our testbed using two UE-powered games.

\subsection{Experiment Setup}
\label{sec:expsetup}

\summary{Testbed.}\quad
\sys\ was evaluated on a testbed consisting of a server and multiple synthetic users.
This controlled environment allows us to manage the number of users and their compression settings, ensuring that \sys's performance can be evaluated and compared to other baselines under consistent conditions.
The testbed runs on a workstation with AMD Ryzen 9 7950X processor with 16 cores and 32 threads, 64 GB main memory, and Nvidia RTX 4090 with 24 GB VRAM.
This machine emulates multiple users by running multiple game instances, each configured with different compression settings to simulate diverse network conditions.

\begin{figure}[h!]
  \centering
  \begin{subfigure}[t]{0.23\textwidth}
    \includegraphics[width=\textwidth]{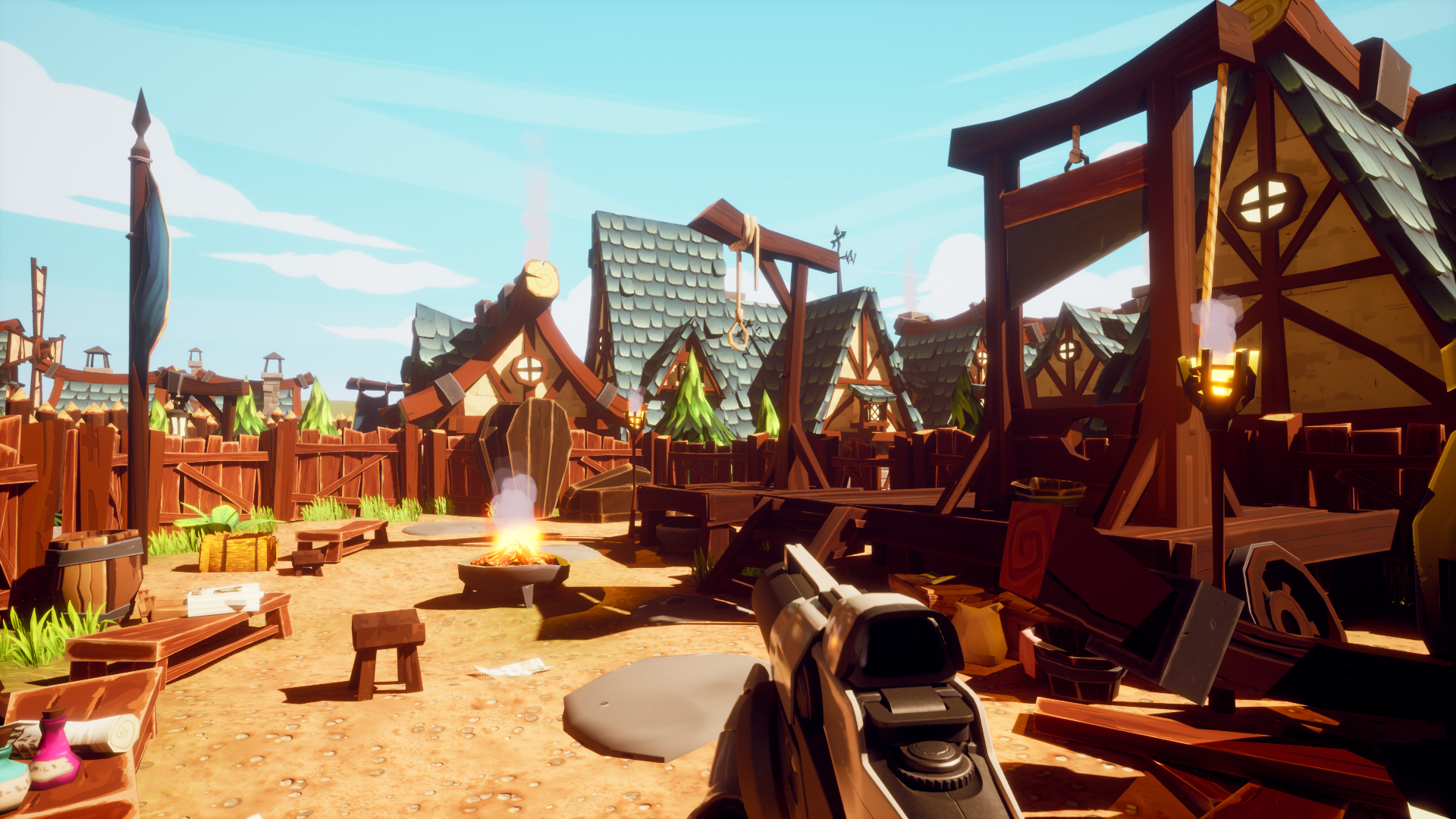}
    \vspace{-3.0ex}
    \caption{Village Shooter}
    \label{fig:exg1}
  \end{subfigure}\hspace{0.01\textwidth}
  \begin{subfigure}[t]{0.23\textwidth}
    \includegraphics[width=\textwidth]{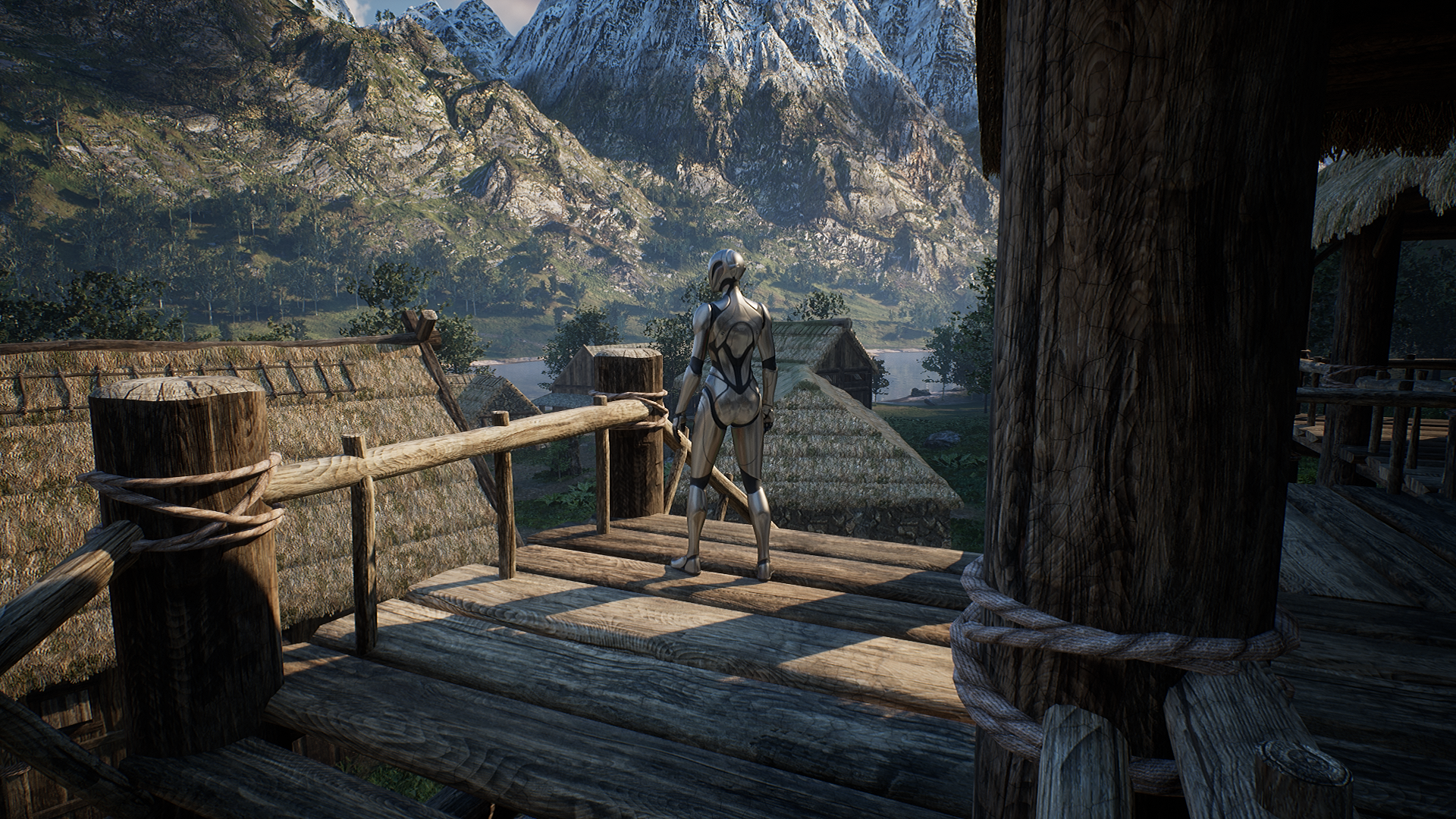}
    \vspace{-3.0ex}
    \caption{Mountain Hiker}
    \label{fig:exg2}
  \end{subfigure}
  \vspace{-1.0ex}
  \caption{The screenshots of sample games}
  \vspace{-2.0ex}
  \label{fig:examplegames}
\end{figure}

\summary{Sample Games.}\quad
Two UE games~\cite{g1,g3}, shown in Figure~\ref{fig:examplegames}, were used for the evaluation.
The games differ in their graphical complexity: Village Shooter features relatively simple, cartoonish graphics, while Mountain Hiker has more complex, realistic graphics, resulting in higher rendering costs.

\summary{Experiment Scenarios.}\quad
We designed three experimental scenarios for evaluation, each involving a server accommodating up to six users divided into two groups (A and B) of three users each.
Within each group, users have different network conditions: Good (G), Fair (F), and Poor (P) with corresponding QPs as outlined in Table~\ref{tab:rqqp_summary}.
The primary distinction between the groups lies in their arrival patterns: Group A users join the server sequentially at 20-second intervals (G at 0 seconds, F at 20 seconds, P at 40 seconds),
whereas all Group B users (G, F, P) join simultaneously at 60 seconds.
These patterns allow us to evaluate \sys's adaptability and performance for both gradual and burst user arrivals.

In Scenario 1 and 2, all users play the same game, Village Shooter and Mountain Hiker, respectively.
In Scenario 3 (Mixed-game Case), Group A plays Village Shooter and Group B plays Mountain Hiker.
This variation in gameplay scenarios helps assess \sys's effectiveness under different rendering workloads.

The RQ optimization interval is set to 5 seconds, with tFPS of 30 and a FPS upper bound of 120.
The cost and quality terms in the efficiency score (Eq.~\ref{eq:score}) are equally weighted (\(\alpha\)=0.5).
For backoff settings, we use a base of 2 with a maximum backoff count of 5.

\subsection{Scalability}
\label{sec:scalability}
We run the experiments in Scenarios 1 and 2 and compare \sys's performance with a naive baseline that naively accommodates more users by assigning \texttt{Very High} RQ to all users.
In this baseline, which is inspired by an existing cloud gaming service~\cite{geforcenow}, each user is assigned a GPU with the highest RQ, and users are queued when all GPUs are occupied.
We use this highest-only case as a proxy to evaluate how such an approach would scale.
The performance is measured by FPS traces and GPU usage.

\begin{figure*}[h!]
  \centering
  \begin{subfigure}[t]{0.49\textwidth}
    \includegraphics[width=\textwidth]{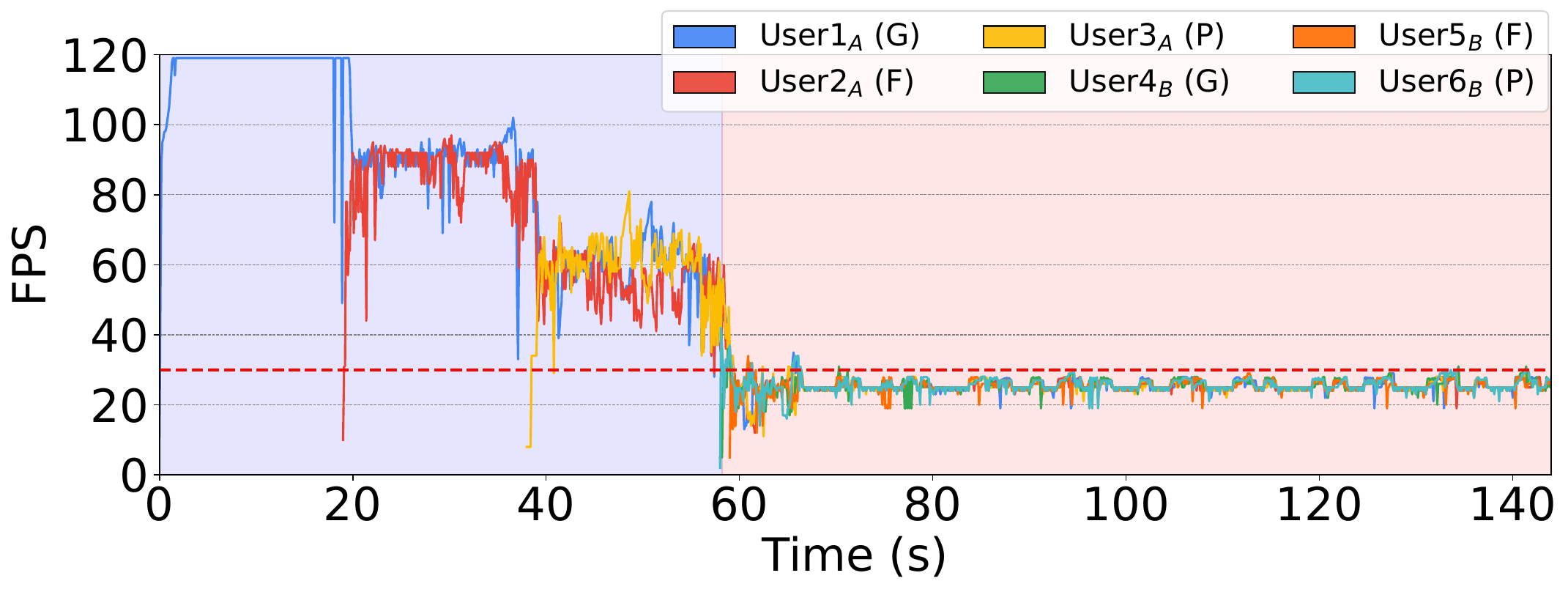}
    \vspace{-4.0ex}
    \caption{The highest-only case: all users with \texttt{Very High} RQ}
    \label{fig:g1_ts_nst}
  \end{subfigure}\hspace{0.01\textwidth}
  \begin{subfigure}[t]{0.49\textwidth}
    \includegraphics[width=\textwidth]{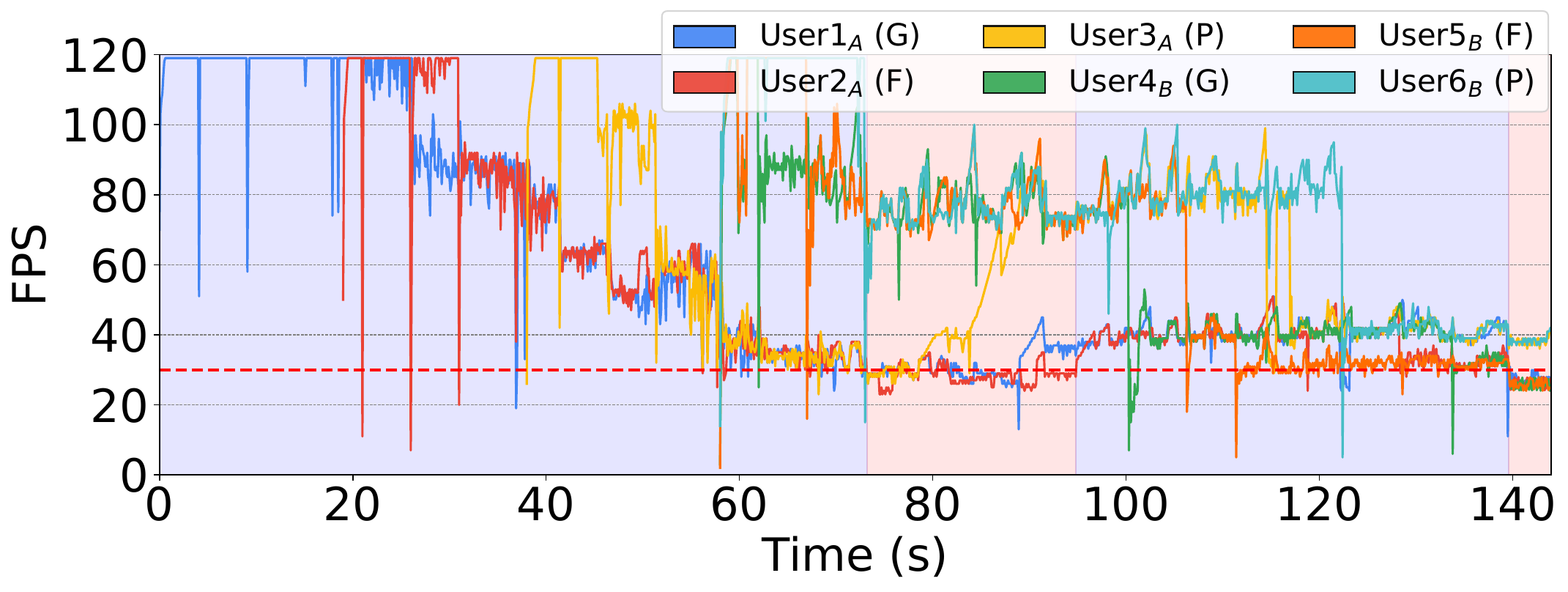}
    \vspace{-4.0ex}
    \caption{\sys: users with adaptive RQ with its optimization}
    \label{fig:g1_ts_st}
  \end{subfigure}
  \vspace{-1.0ex}
  \caption{The FPS traces of Village Shooter with and without \sys. The dashed line is the FPS threshold (30),
  and the blue background color indicates all users are served with higher FPS than the threshold, while the red indicates the opposite.}
  \vspace{-1.0ex}
  \label{fig:g1_ts}
\end{figure*}

\begin{figure*}[h!]
  \centering
  \begin{subfigure}[t]{0.49\textwidth}
    \includegraphics[width=\textwidth]{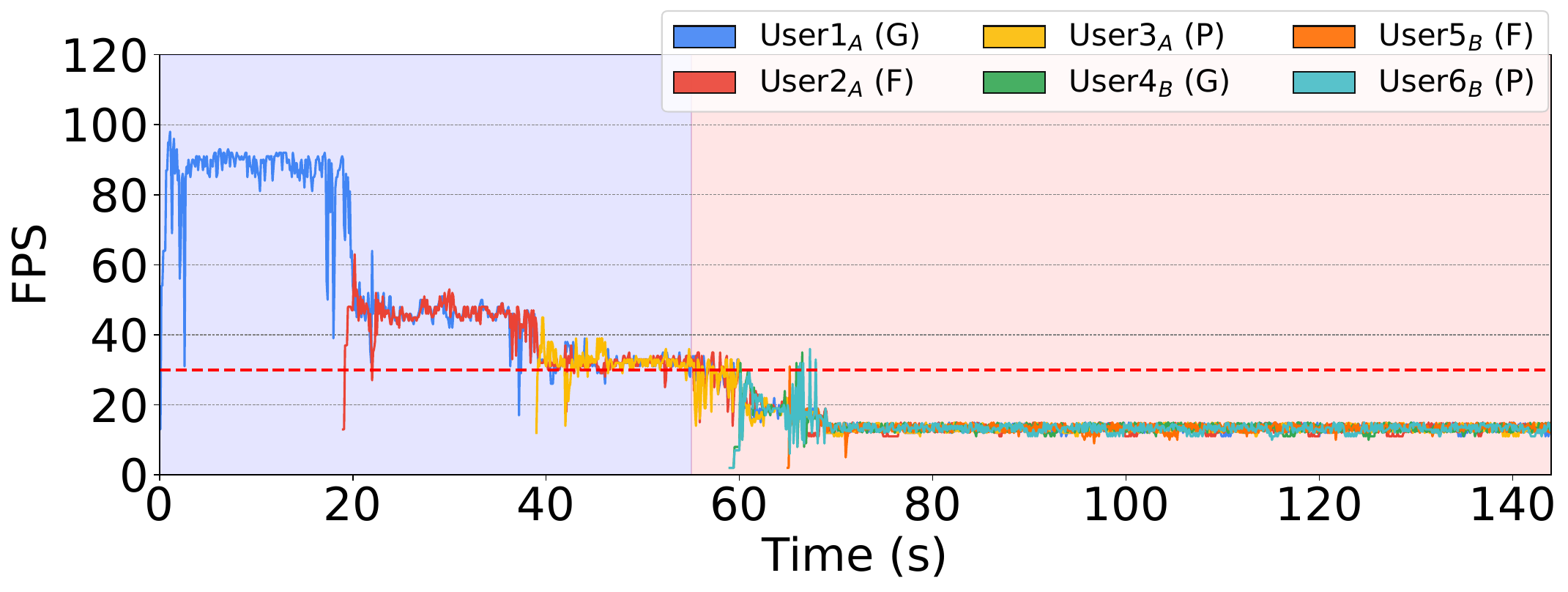}
    \vspace{-4.0ex}
    \caption{The highest-only case: all users with \texttt{Very High} RQ}
    \label{fig:g2_ts_nst}
  \end{subfigure}\hspace{0.01\textwidth}
  \begin{subfigure}[t]{0.49\textwidth}
    \includegraphics[width=\textwidth]{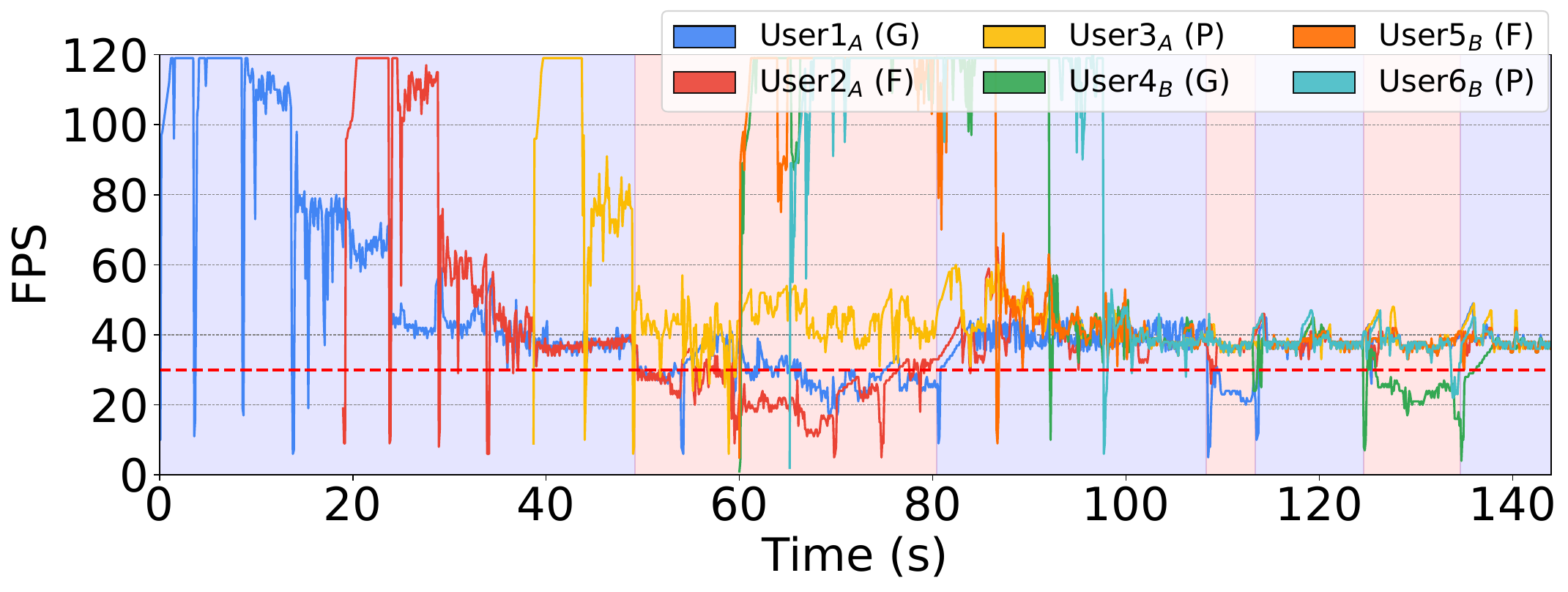}
    \vspace{-4.0ex}
    \caption{\sys: users with adaptive RQ with its optimization}
    \label{fig:g2_ts_st}
  \end{subfigure}
  \vspace{-1.0ex}
  \caption{The FPS traces of Mountain Hiker with and without \sys. The dashed line is the FPS threshold (30),
  and the blue background color indicates all users are served with higher FPS than the threshold, while the red indicates the opposite.}
  \vspace{-1.0ex}
  \label{fig:g2_ts}
\end{figure*}

\begin{figure*}[h!]
  \centering
  \begin{subfigure}[t]{0.49\textwidth}
    \includegraphics[width=\textwidth]{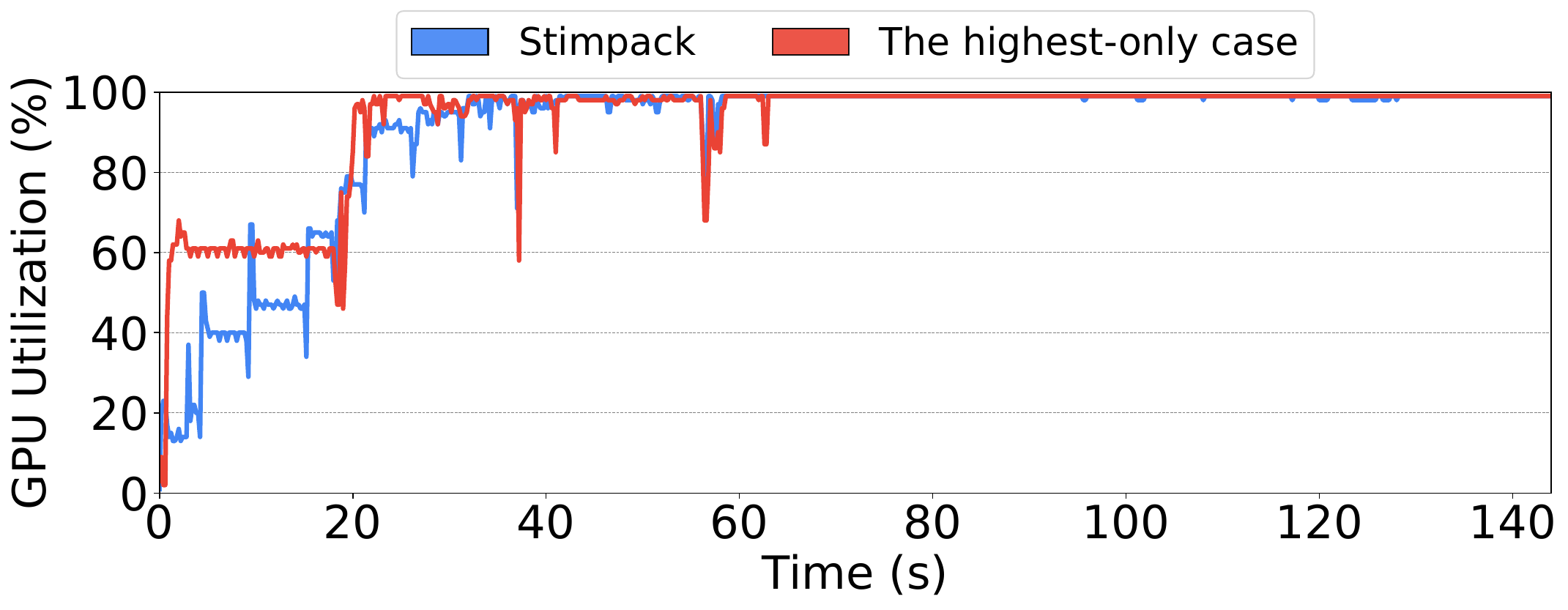}
    \vspace{-4.0ex}
    \caption{GPU usage with Village Shooter (matched to Figure~\ref{fig:g1_ts})}
    \label{fig:gpu_g1}
  \end{subfigure}\hspace{0.01\textwidth}
  \begin{subfigure}[t]{0.49\textwidth}
    \includegraphics[width=\textwidth]{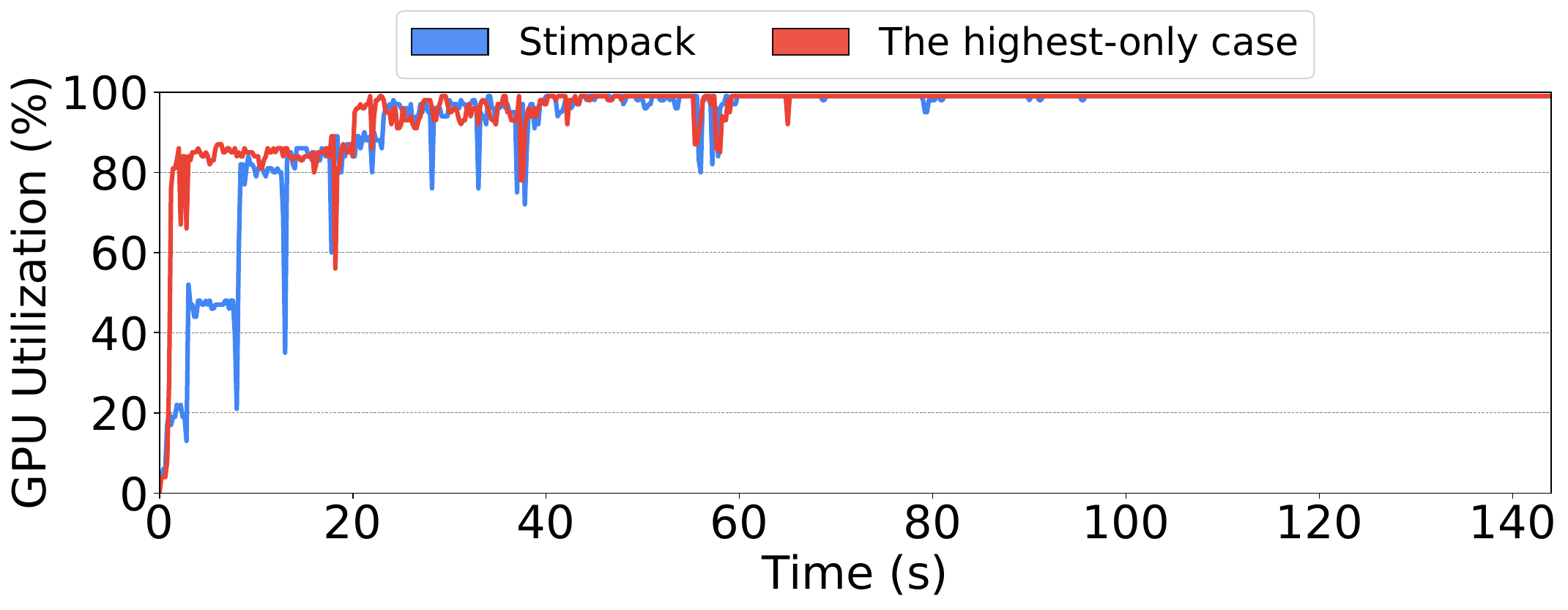}
    \vspace{-4.0ex}
    \caption{GPU usage with Mountain Hiker (matched to Figure~\ref{fig:g2_ts})}
    \label{fig:gpu_g2}
  \end{subfigure}
  \vspace{-1.0ex}
  \caption{The server's GPU usage of \sys\ and the highest-only case}
  \vspace{-2.0ex}
  \label{fig:gpu_usage}
\end{figure*}

\subsubsection{Village Shooter: FPS Trace}
Figure~\ref{fig:g1_ts} shows the FPS trace graphs of Village Shooter.

\summary{Highest-only Case.}\quad
For the initial 3 users in Group A, User1$_A$ (G) is served with 120 FPS and \texttt{Very High} RQ (Figure~\ref{fig:g1_ts_nst}).
When User2$_A$ (F) joins the server, both users are served with \tlide93 FPS due to the increased load (as described with Figure~\ref{fig:vis}).
users' FPS further drops to \tlide60 FPS when User3$_A$ (P) arrives.
When Group B joins the server simultaneously, the FPS for all users drops to \tlide28 FPS, falling below tFPS.

\summary{\sys.}\quad
User1$_A$ (G) begins with a \texttt{Low} RQ setting, as \sys\ initially assigns the lowest rendering quality to users.
Due to the high FPS achieved, \sys\ progressively promotes User1$_A$'s RQ.
As shown in Figure~\ref{fig:g1_ts_st}, User1$_A$ (G) eventually reaches the same state as in the highest-only case, achieving \texttt{Very High} RQ and 120 FPS after RQ promotions at \tlide18 seconds.
When User2$_A$ (F) and User3$_A$ (P) join the server, they also start with \texttt{Low} RQ and maintain higher FPS compared to the highest-only case.
As all three users consistently achieve above tFPS, \sys\ continues to promote their RQs, enabling them to reach the same state as the highest-only case by around 50 seconds.

At the 60-second mark, Group B joins the server with \texttt{Low} RQ, and Group A still maintains FPS above tFPS.
\sys\ then begins promoting the RQs of Group B.
User4$_B$ (G) and User5$_B$ (F) are promoted to \texttt{Medium} RQ first, as these promotions are expected to be more effective in improving visual quality with their better network conditions.
After these promotions, Group A, which is at \texttt{Very High} RQ, sees its FPS drop to \tlide35 FPS, while User4$_B$ and User5$_B$ reach \tlide80 FPS with \texttt{Medium}.
When User6$_B$ (P) is promoted around 70 seconds, FPS for \texttt{Very High} RQ users drop below tFPS, as indicated by the red background in Figure~\ref{fig:g1_ts_st}.

To address these users with below tFPS, \sys\ initiates a demotion phase.
During this phase, User3$_A$ (P) is initially demoted to \texttt{High}, followed by further demotions of the remaining users at \texttt{Very High}.
Around 95 seconds, these adjustments make all users' FPS above tFPS.
\sys\ then continues to optimize users' RQs based on their \(Score_u\), balancing FPS and user-perceived visual quality.
In the later part of the trace in Figure~\ref{fig:g1_ts_st}, \sys\ converges users' RQs to optimized levels, as summarized in Table~\ref{tab:scal}.

\subsubsection{Mountain Hiker: FPS Trace}
Figure~\ref{fig:g2_ts} illustrates the FPS traces of Mountain Hiker, a game with higher rendering resource demands.

\summary{Highest-only case.}\quad
In the highest-only case (Figure~\ref{fig:g2_ts_nst}), User1$_A$ (G) initially experiences \tlide94 FPS.
As User2$_A$ and User3$_A$ join the server, FPS of Group A decreases to \tlide32.
When Group B joins the server, it becomes overloaded, and the FPS for all users plummets to \tlide13 FPS, which is too low to provide playable gaming experiences.

\summary{\sys.}\quad
The effectiveness of \sys\ is more pronounced with Mountain Hiker's intensive workload.
Initially, User1$_A$ (G) and User2$_A$ (F) start with \texttt{Low} RQ and are subsequently promoted to \texttt{Very High}, reaching the same state as the highest-only case.
However, when User3$_A$ (P) joins and is promoted to \texttt{High} RQ around 50 seconds, the FPS of User1$_A$ and User2$_A$ falls below tFPS.
This drop is attributed to the overhead of RQ adjustment, as demonstrated in \S\ref{sec:rq_oscillation}; the highest-only case serves Group A with near tFPS without this overhead.

The situation further deteriorates when Group B joins at 60 seconds.
Despite starting with \texttt{Low} RQ, the server is already heavily loaded and struggles to maintain users' FPS above tFPS.
\sys\ responds by keeping Group B at \texttt{Low} and demoting the RQs of the other users.
User3$_A$ (P) is demoted continuously to \texttt{Medium} because the quality loss from demoting a user with a poor network condition is expected to be less significant.
However, User1$_A$ and User2$_A$ are also demoted to \texttt{Medium} as their FPS remains below tFPS.
The FPS of all users becomes higher than tFPS around 80 seconds.

After ensuring the FPS of all users is above tFPS, \sys\ begins promoting users' RQs in subsequent optimization rounds.
Group B users are promoted from \texttt{Low} to \texttt{Medium} RQ, resulting in \tlide37 FPS for all users at \texttt{Medium}.
\sys\ then attempts to promote User1$_A$ (G) around 110 seconds, but this causes User1$_A$'s FPS to drop below tFPS, leading to a demotion around 115 seconds.
Detecting this oscillation, \sys's backoff mechanism is activated to prevent further oscillatory RQ adjustments.
As the trace progresses, the backoff duration increases, and users' RQs stabilize with FPS above tFPS, as summarized in Table~\ref{tab:scal}.

\begin{table*}[ht!]
  \caption{\label{tab:scal} The summary of the user serving states of \sys\ and the other baselines}
  \vspace{-2.0ex}
  \scriptsize
  \centering
  \begin{tabularx}{0.91\textwidth}{|X||c|c|c|c|c|c|c|c|c|c|}
    \cline{1-11}
    \multicolumn{2}{|c|}{\multirow{2}{*}{}}& \multicolumn{3}{c|}{Village Shooter (Scenario 1)} & \multicolumn{3}{c|}{Mountain Hiker (Scenario 2)} & \multicolumn{3}{c|}{\makecell{Mixed-game Case (Scenario 3)\\($A$ for Village Shooter, $B$ for Mountain Hiker)}}      \\ \cline{3-11}
    \multicolumn{2}{|c|}{\multirow{2}{*}{}}       & RQ                 & FPS  & \makecell{Visual\\Quality}  & RQ & FPS & \makecell{Visual\\Quality}     & RQ & FPS & \makecell{Visual\\Quality} \\ \hhline{===========}
    \multirow{6}{*}{\textbf{\sys}}     & User1$_A$ (G) & \texttt{Very High} & 30.88 & \VHG          & \texttt{Medium}   & 37.91  & \MG            & \texttt{Very High}  & 30.17    & \VHG \\\cline{2-11}
                                       & User2$_A$ (F) & \texttt{High}      & 37.91 & \HM           & \texttt{Medium}   & 37.74  & \MM            & \texttt{High}       & 38.64    & \HM  \\\cline{2-11}
                                       & User3$_A$ (P) & \texttt{Medium}    & 72.05 & \MP           & \texttt{Medium}   & 35.4   & \MP            & \texttt{Medium}     & 61.05    & \MP  \\\cline{2-11}
                                       & User4$_B$ (G) & \texttt{Very High} & 30.41 & \VHG          & \texttt{Medium}   & 37.81  & \MG            & \texttt{Medium}     & 38.64    & \MG  \\\cline{2-11}
                                       & User5$_B$ (F) & \texttt{High}      & 39.22 & \HM           & \texttt{Medium}   & 36.85  & \MM            & \texttt{Medium}     & 38.33    & \MM  \\\cline{2-11}
                                       & User6$_B$ (P) & \texttt{Medium}    & 69.54 & \MP           & \texttt{Medium}   & 35.87  & \MP            & \texttt{Medium}     & 37.7     & \MP  \\\hhline{===========}

    \multirow{6}{*}{Highest-only} & User1$_A$ (G) & \multirow{6}{*}{\texttt{Very High}} & 27.69 & \VHG  & \multirow{6}{*}{\texttt{Very High}}  & 13.4  & \VHG & \multirow{6}{*}{\texttt{Very High}} & 25.32 & \VHG \\\cline{2-2}\cline{4-5}\cline{7-8}\cline{10-11}
                                  & User2$_A$ (F) &                                     & 28.57 & \VHM  &                                      & 13.51 & \VHM &                                     & 25.64 & \VHM \\\cline{2-2}\cline{4-5}\cline{7-8}\cline{10-11}
                                  & User3$_A$ (P) &                                     & 27.87 & \VHP  &                                      & 13.09 & \VHP &                                     & 26.06 & \VHP \\\cline{2-2}\cline{4-5}\cline{7-8}\cline{10-11}
                                  & User4$_B$ (G) &                                     & 28.07 & \VHG  &                                      & 13.25 & \VHG &                                     & 14.36 & \VHG \\\cline{2-2}\cline{4-5}\cline{7-8}\cline{10-11}
                                  & User5$_B$ (F) &                                     & 28.78 & \VHM  &                                      & 13.75 & \VHM &                                     & 13.87 & \VHM \\\cline{2-2}\cline{4-5}\cline{7-8}\cline{10-11}
                                  & User6$_B$ (P) &                                     & 28.47 & \VHP  &                                      & 13.46 & \VHP &                                     & 14.01 & \VHP \\\hhline{===========}

    \multirow{6}{*}{Lowest-only}  & User1$_A$ (G) & \multirow{6}{*}{\texttt{Low}} & \multirow{6}{*}{120} & \LG & \multirow{6}{*}{\texttt{Low}} & \multirow{6}{*}{120} & \LG & \multirow{6}{*}{\texttt{Low}} & \multirow{6}{*}{120} & \LG \\\cline{2-2}\cline{5-5}\cline{8-8}\cline{11-11}
                                  & User2$_A$ (F) &                               &                      & \LM &                               &                      & \LM &                               &                      & \LM \\\cline{2-2}\cline{5-5}\cline{8-8}\cline{11-11}
                                  & User3$_A$ (P) &                               &                      & \LP &                               &                      & \LP &                               &                      & \LP \\\cline{2-2}\cline{5-5}\cline{8-8}\cline{11-11}
                                  & User4$_B$ (G) &                               &                      & \LG &                               &                      & \LG &                               &                      & \LG \\\cline{2-2}\cline{5-5}\cline{8-8}\cline{11-11}
                                  & User5$_B$ (F) &                               &                      & \LM &                               &                      & \LM &                               &                      & \LM \\\cline{2-2}\cline{5-5}\cline{8-8}\cline{11-11}
                                  & User6$_B$ (P) &                               &                      & \LP &                               &                      & \LP &                               &                      & \LP \\\hhline{===========}

    \multirow{12}{*}{dJay~\cite{grizan2015djay}}
     & User1$_A$ (G) &                                                 & \makecell{39.35\\(25.81)} & \makecell{\HG\\(\VHG)}
                      &                                                 & \makecell{34.56\\(19.9)}  & \makecell{\MG\\(\HG)}
                      &                                                 & \makecell{37.9 \\(23.44)} & \makecell{\HG\\(\VHG)} \\\cline{2-2}\cline{4-5}\cline{7-8}\cline{10-11}

     & User2$_A$ (F) &                                                 & \makecell{40.61\\(26.1)}  & \makecell{\HM\\(\VHM)}
                      &                                                 & \makecell{34.41\\(19.5)}  & \makecell{\MM\\(\HM)}
                      &  \makecell{\texttt{High}\\(\texttt{Very High})} & \makecell{37.38\\(23.7)}  & \makecell{\HM\\(\VHM)}\\\cline{2-2}\cline{4-5}\cline{7-8}\cline{10-11}

     & User3$_A$ (P) & \texttt{High}                                   & \makecell{40.48\\(26.2)}  & \makecell{\HP\\(\VHP)}
                      & \texttt{Medium}                                 & \makecell{36.44\\(19.78)} & \makecell{\MP\\(\HP)}
                      &                                                 & \makecell{37.4 \\(23.81)} & \makecell{\HP\\(\VHP)} \\\cline{2-2}\cline{4-5}\cline{7-11}

     & User4$_B$ (G) & (\texttt{Very High})                            & \makecell{39.63\\(26.41)} & \makecell{\HG\\(\VHG)}
                      & (\texttt{High})                                 & \makecell{37.11\\(19.34)} & \makecell{\MG\\(\HG)}
                      &                                                 & \makecell{36.59\\(13.4)}  & \makecell{\MG\\(\HG)} \\\cline{2-2}\cline{4-5}\cline{7-8}\cline{10-11}

     & User5$_B$ (F) &                                                 & \makecell{40.01\\(25.9)}  & \makecell{\HM\\(\VHM)}
                      &                                                 & \makecell{36.8\\(19.48)}  & \makecell{\MM\\(\HM)}
                      &  \makecell{\texttt{Medium}\\(\texttt{High})}    & \makecell{37.08\\(13.21)} & \makecell{\MM\\(\HM)} \\\cline{2-2}\cline{4-5}\cline{7-8}\cline{10-11}

     & User6$_B$ (P) &                                                 & \makecell{39.93\\(26.21)} & \makecell{\HP\\(\VHP)}
                      &                                                 & \makecell{37.2\\(19.14)}  & \makecell{\MP\\(\HP)}
                      &                                                 & \makecell{36.9 \\(12.97)} & \makecell{\MP\\(\HP)} \\\cline{1-11}
  \end{tabularx}
  \vspace{-3.0ex}
\end{table*}

\subsubsection{Summary}
Our scalability evaluation demonstrates that \sys\ significantly outperforms a naive baseline.
While the highest-only approach suffers a severe performance collapse that renders the games unplayable under increased user load, \sys\ maintains a playable FPS for all users.
Specifically, in the highest-only case, the maximum number of users that can be served with an above-threshold FPS is 5 for Village Shooter and 3 for Mountain Hiker, as shown in Figures~\ref{fig:g1_ts_nst} and~\ref{fig:g2_ts_nst}.
\sys, on the other hand, improves the user capacity of the server by 1.2\mul\ for Village Shooter and 2\mul\ for Mountain Hiker, while using the same resource footprint (Figure~\ref{fig:gpu_usage}).
This proves that our approach effectively addresses the scalability bottleneck on resource-constrained servers by adaptively adjusting users' RQ, which allows each GPU to accommodate a more number of users.

\subsection{Gaming Service Quality}
\label{sec:servicequality}
Having demonstrated \sys's ability to improve server scalability, we now evaluate how well it maintains gaming service quality and user experience compared to various baselines.
This section addresses the second key question of our evaluation: whether \sys\ provides a better user experience beyond simply accommodating more users per GPU.
To provide a comprehensive analysis, we perform both a quantitative comparison using a service quality metric and a qualitative comparison through a user study.


\subsubsection{Baselines}
Along with the highest-only case, we consider two other baselines: the lowest-only case and dJay~\cite{grizan2015djay}.
The lowest-only case represents a naive scenario where all users are assigned the lowest RQ.
While the highest-only case maximizes visual quality, the lowest-only case prioritizes user accommodation with high FPS, offering contrasting perspectives on resource allocation strategies.

dJay is a previous work that adjusts users' RQs to reduce server load.
Although it shares similarities with \sys\ in terms of RQ adjustment, there are significant differences in approach.
Firstly, dJay adapts RQs to maintain FPS but does not consider the user-perceived visual quality affected by compression settings under varying network conditions.
Secondly, while dJay is also round-based, it simultaneously adjusts all users' RQs every round without a stabilization mechanism.

In contrast, \sys\ addresses these limitations with a holistic approach.
It intelligently selects the most efficient RQ adjustments among users based on server-side rendering costs and the estimated user-perceived visual quality, which is affected by their QPs.
This prioritization is enabled by \(Score_u\) (\S\ref{sec:scoring}) with our user-side quality estimation model (\S\ref{sec:frameprediction}).
Additionally, \sys\ incorporates a stabilization mechanism to mitigate the negative effects caused by RQ oscillation overheads (\S\ref{sec:rq_oscillation}).
These features allow \sys\ to optimize performance for multi-user scenarios, effectively managing the trade-off between resource efficiency and service quality in ways that existing baselines cannot.
The highest-only, lowest-only, and dJay cases thus serve as reference points for our evaluation.

\subsubsection{Quantitative Comparison}
\label{sec:servicequality_res}
\summary{Service Quality Metric.}\quad
Gaming service quality assessment requires the comprehensive consideration of both FPS and visual quality.
To our best knowledge, no established single metric holistically encompasses both factors.
While FPS and visual quality have traditionally been measured and used separately to estimate gaming experience, we introduce an aggregate metric - the service quality score (Eq.~\ref{eq:sqs}) - to facilitate a direct comparison with the baselines.
This score is calculated as the product of the FPS score (Eq.~\ref{eq:fps_score}), which is designed to reflect human visual perception, and the visual quality measured by VMAF, with a higher score indicating better performance.
We have further discussion on the metric design and need of comprehensive metric in Appendix~\ref{sec:metric_validity}.

\vspace{-1.5ex}
\begin{equation}\label{eq:sqs}
   Service\ Quality\ Score = FPS\_Score_u \cdot VMAF_u
\end{equation}
\vspace{-1.5ex}

\begin{figure*}[h!]
  \centering
  \begin{subfigure}[t]{0.32\textwidth}
    \includegraphics[width=\textwidth]{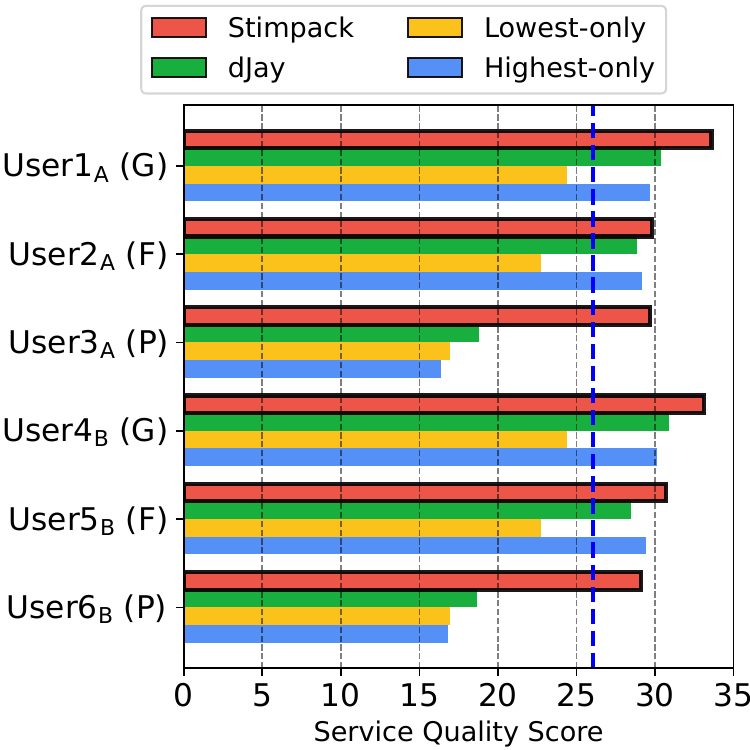}
    \caption{Scenario 1: Village Shooter}
    \label{fig:g1_sq}
  \end{subfigure}\hspace{0.01\textwidth}
  \begin{subfigure}[t]{0.32\textwidth}
    \includegraphics[width=\textwidth]{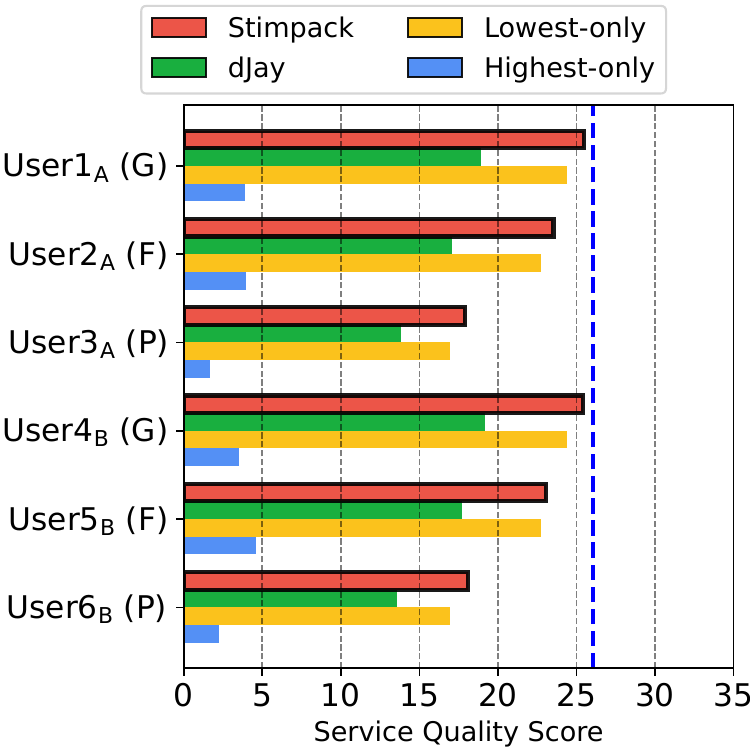}
    \caption{Scenario 2: Mountain Hiker}
    \label{fig:g2_sq}
  \end{subfigure}\hspace{0.01\textwidth}
  \begin{subfigure}[t]{0.32\textwidth}
    \includegraphics[width=\textwidth]{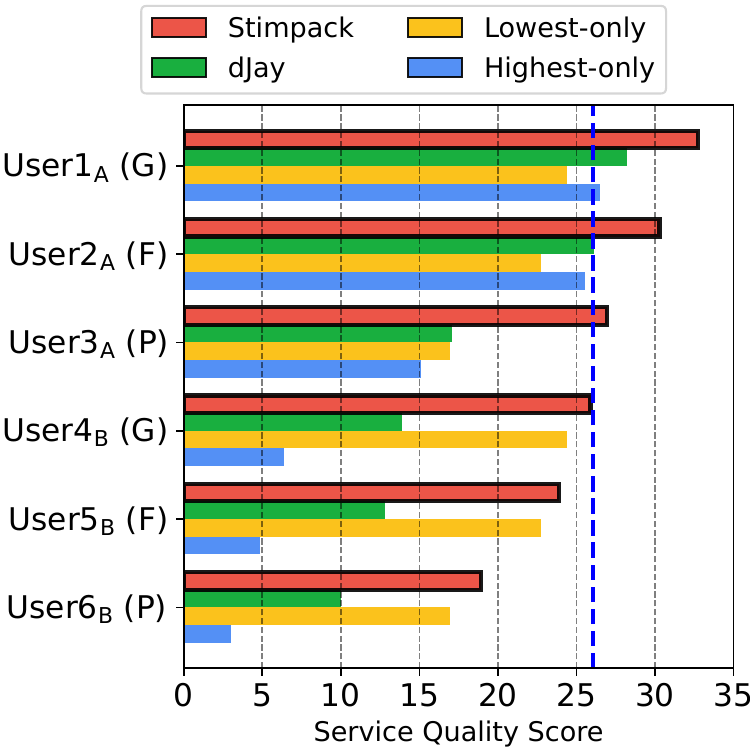}
    \caption{Scenario 3: Mixed games}
    \label{fig:mixed_sq}
  \end{subfigure}
  \vspace{-1.5ex}
  \caption{Service quality score comparison of \sys\ and the other baselines. The dashed line corresponds to the score of 30 FPS with \texttt{High} RQ and good network condition.}
  \label{fig:sq_res}
  \vspace{-3.0ex}
\end{figure*}

\summary{Service Quality Comparison.}\quad
Table~\ref{tab:scal} summarizes the serving states of users under \sys\ and the baselines across three experimental scenarios: Scenario 1 (Village Shooter), Scenario 2 (Mountain Hiker), and Scenario 3 (Mixed-game Case).
In Scenario 3, Group A plays Village Shooter, while Group B plays Mountain Hiker.
Since dJay lacks a stabilization mechanism, it faces RQ oscillation overheads.
Therefore, we report FPS and visual quality metrics for the two oscillating RQ levels (shown in parentheses) that it alternates between.
\sys's results, by contrast, are the final stabilized serving states.

Figure~\ref{fig:sq_res} shows the comparative results of the service quality scores based on the serving states in Table~\ref{tab:scal}.
dJay's score is the average of the two oscillating RQ levels.

\summary{Scenario 1.}\quad
In Figure~\ref{fig:g1_sq}, \sys\ (red) shows the better service quality scores across all users.
For users with (G) and (F) network conditions, the highest-only case (blue) achieves comparable results to \sys, as their FPS remains close to, though slightly below, tFPS.
However, the highest-only case shows quality degradation for User3$_A$ and User6$_B$ (P) due to its failure to account for compression losses under network conditions, which results in inefficient resource usage.

dJay's performance (green) falls short of \sys\ due to RQ oscillation overheads for users with (G) and (F) network conditions.
Similar to the highest-only case, dJay's inability to account for varying network conditions leads to inefficient resource usage for User3$_A$ and User6$_B$ (P).
At the other extreme, the lowest-only case (yellow), despite high FPS, delivers consistently poor service quality due to its minimal RQ that results in low visual quality.

\sys\ achieves better service quality through optimized resource allocation, enhancing visual quality for users with (G) and (F) network conditions while reducing RQs for users with (P) to prevent resource waste.
In this scenario, the most competitive baseline is dJay.
\sys\ outperforms dJay by 19\% in the averaged score across all users: 31.2 for \sys\ and 26.2 for dJay.

\summary{Scenario 2.}\quad
In Figure~\ref{fig:g2_sq}, the service quality results differ notably due to Mountain Hiker's higher resource demands, resulting in lower service quality scores across all cases.
The highest-only case shows particularly poor results as its FPS drops far below the playable threshold (\tlide13 FPS).

For dJay, the performance difference compared to \sys\ becomes less pronounced for User3$_A$ and User6$_B$ (P).
This is because the heavy load limits both systems' ability to increase RQ, thereby reducing dJay’s tendency to waste resources with RQ adjustments that are agnostic to user-side quality.

In Scenario 2, the lowest-only case is the most competitive baseline, as it maintains smooth gameplay.
\sys\ outperforms it by a small margin, 5\% in the averaged score across all users: 22.4 for \sys\ and 21.3 for the lowest-only case.
As summarized in Table~\ref{tab:scal}, \sys\ optimizes performance by stabilizing all users' RQ at \texttt{Medium}, striking a balance between maintaining FPS and visual quality.
The serving states of \sys\ (\texttt{Medium}, 38 FPS) achieve better results with the aggregate metric than those of the lowest-only case (\texttt{Low}, 120 FPS).
However, users may have different subjective preferences; some may prioritize visual quality over FPS, while others may prefer smoother gameplay.
We further explore the user preferences in our qualitative user study.

\summary{Scenario 3.}\quad
In Figure~\ref{fig:mixed_sq}, the mixed-game scenario shows \sys's versatility in handling heterogeneous workloads.
For Group A playing Village Shooter (moderate resource demands), \sys\ maintains higher RQs for User1$_A$ (G) and User2$_A$ (F) similar to the highest-only case.
Unlike the highest-only case, \sys\ adjusts the RQ of User3$_A$ (P) to an efficient level, balancing visual quality and FPS.
For Group B playing Mountain Hiker (higher resource demands), the efficient RQ levels are close to the lowest-only case.
In this scenario, \sys\ achieves 24\% higher service quality than the lowest-only case, with scores of 26.5 and 21.3, respectively.

The results demonstrate \sys's adaptability in finding efficient operating states between the two extremes.
When a game's resource demand is high and a server is heavily loaded, \sys\ may converge to the lowest RQ settings to maintain above-threshold FPS, similar to the lowest-only case.
Conversely, when resources are sufficient and a game's resource demand is low, \sys\ matches the highest RQ settings to improve visual quality with a playable FPS.
Through its user-side quality-aware RQ optimization, \sys\ sets efficient RQ settings for users of different network conditions based on game workloads and serving states, maximizing aggregate service quality.

\subsubsection{Qualitative User Evaluation}
\label{sec:userstudy}
\summary{Methodology.}\quad
To validate \sys's performance, we conduct a user study based on the mixed-game case (Scenario 3).
The study involves 30 participants aged 20-58, with varying levels of video game familiarity: 11 very familiar, 15 somewhat familiar, and 4 not really familiar.

Participants compare pairs of anonymous recorded gaming clips (A and B), where one is served by \sys\ and the other by a baseline.
They are asked to choose which clip they preferred in terms of smoothness and visual quality, or indicate a draw, following a subjective evaluation methodology~\cite{ITUTP910}.
Following the initial survey, we asked participants to provide additional feedback to explore the reasons behind their choices.
Based on the responses, a winning rate is calculated using the formula in Eq.~\ref{eq:wr}, which serves as a metric to compare \sys's performance against the baselines.

\begin{equation}\label{eq:wr}
  Winning\ Rate = \frac{Wins + 0.5\cdot Draws}{Total\ Comparisons}
\end{equation}

\begin{figure}[t]
  \centering
  \begin{subfigure}[t]{0.23\textwidth}
    \includegraphics[width=\textwidth]{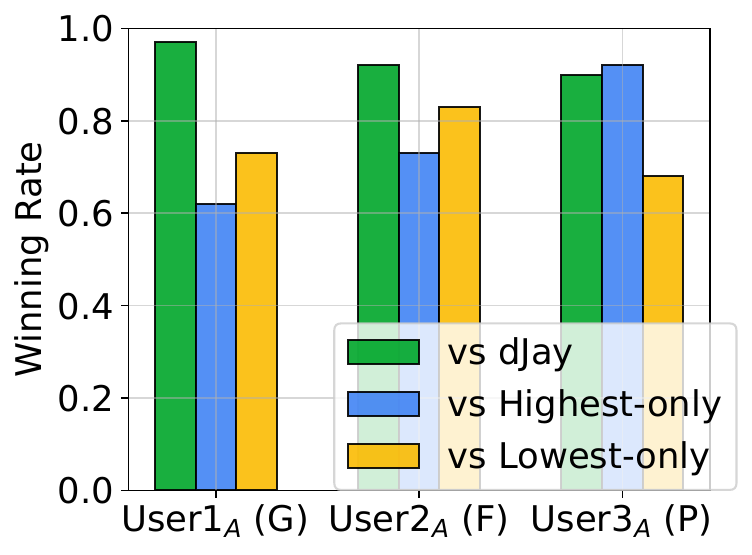}
    \vspace{-3.0ex}
    \caption{Village Shooter}
    \label{fig:us1}
  \end{subfigure}\hspace{0.01\textwidth}
  \begin{subfigure}[t]{0.23\textwidth}
    \includegraphics[width=\textwidth]{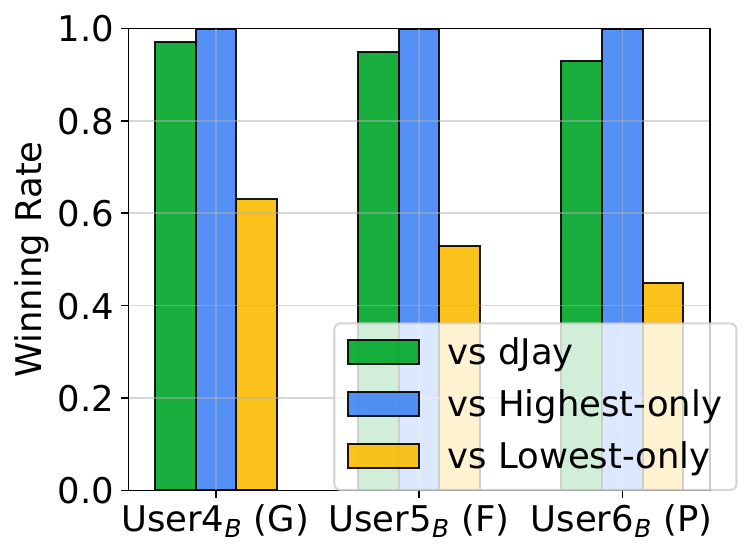}
    \vspace{-3.0ex}
    \caption{Mountain Hiker}
    \label{fig:us2}
  \end{subfigure}
  \vspace{-1.0ex}
  \caption{The winning rate of \sys\ against the other baselines from the user study under Scenario 3}
  \vspace{-2.0ex}
  \label{fig:us}
\end{figure}

\summary{\sys\ vs dJay.}\quad
As shown in Figure~\ref{fig:us}, \sys\ consistently outperforms dJay across all users.
This demonstrates the effectiveness of \sys's holistic approach, which incorporates both user-side quality-aware RQ adaptation and a stabilization mechanism.
One notable observation from our user feedback is that the unstable gaming experience caused by dJay's lack of a stabilization mechanism was also a factor in users' preferences for \sys.
While this aspect was not directly captured in the quantitative score results, it underscores the importance of smooth and stable gaming experiences, which \sys\ delivers effectively.

\summary{\sys\ vs Highest-only.}\quad
For Mountain Hiker, \sys\ significantly outperforms the highest-only case, since the latter struggles to maintain playable FPS.
For Village Shooter, \sys's winning rate decreases as a user's network condition gets better: 92\% for User3$_A$ (P), 73\% for User2$_A$ (F), and 62\% for User1$_A$ (G).
These results align with the service quality scores, where the highest-only case becomes more competitive as the benefits of its higher RQ settings become pronounced for users with better network conditions and FPS slightly below tFPS.

\summary{\sys\ vs Lowest-only.}\quad
For Village Shooter, \sys\ shows a high winning rate against the lowest-only case.
Interestingly, \sys's winning rate for User2$_A$ (F) is 83\%, which is higher than the 73\% for User1$_A$ (G).
Based on the feedback, this result is attributed to subjective preferences with different weightings between visual quality and smoothness.
User1$_A$ (G) is served with (\texttt{Very High}, \tlide30 FPS) by \sys.
Compared to the lowest-only case (\texttt{Low}, 120 FPS), some participants preferred the smoother clip, while others favored the better visual quality.
As User2$_A$ (F) is served with (\texttt{High}, 39 FPS), the participants perceived a more balanced experience with both good visual quality and smoothness, leading to a higher winning rate than User1$_A$ (G).


For Mountain Hiker, \sys's winning rate is lower than the lowest-only case except for the good network condition: 56\% for User4$_B$ (G), 47\% for User5$_B$ (F), and 35\% for User6$_B$ (P).
\sys\ optimizes for \texttt{Medium} RQ with \tlide38 FPS for Group B, while the lowest-only case provides \texttt{Low} RQ with 120 FPS.
Participants' feedback suggests that in this situation, many users prefer the significant smoothness of the lowest-only case over the visual quality provided by \sys.
This finding highlights a more fundamental issue: a lack of an objective metric for gaming experience that comprehensively captures a user's subjective trade-off between FPS and visual quality.
The development of such a metric would enable better system performance, but we leave this as future work as it is beyond the scope of this paper.

\subsubsection{Summary}
Our evaluation on service quality demonstrates that \sys's core mechanisms effectively manage service quality, going beyond simply serving more users.
Through quantitative evaluation, we show that in both gradual and bursty scenarios, \sys\ utilizes the same resources to accommodate more users by adapting and stabilizing their RQs, achieving up to 24\% higher service quality scores compared to the most competitive baseline in each scenario.
Additionally, we conducted further evaluations under the situation where users' network conditions fluctuate and demonstrated that \sys\ effectively operates with such real-world network dynamics (Appendix~\ref{sec:app_dynamics_stress}).
From a qualitative perspective, while user gaming experience is subjective and further research is needed to develop a comprehensive metric that captures this subjectivity, our user study shows \sys's features are effective in maintaining service quality in most scenarios with a high winning rate against the baselines.

\section{Related Work}
\label{sec:relatedwork}
Previous research has focused on improving cloud gaming service quality and efficiency through various approaches.
One popular method involves optimizing server provisioning and allocation based on user demand and geographical distribution to minimize network latency and service operational costs~\cite{deng2016server,deng2017server,basiri2016delay,gao2019cost,slivar2019qoe,yami2020sara,li2020towards}.

Another line of research has explored enhancing server resource efficiency through GPU scheduling and resource allocation, utilizing virtual machines (VMs) and virtualized GPUs~\cite{herrera2014nvidia,qi2014vgris,guan2014energy,zhang2013vgasa,zhang2015cloud,zhang2017fine}.
From the perspective that \sys\ utilizes an application-level knob to adapt user workloads and improve per-GPU scalability, it is orthogonal to those previous works focused on server allocation and GPU scheduling.
This suggests potential opportunities for joint optimizations with previously proposed techniques to reduce the operational cost of cloud gaming services while maintaining service quality and availability.

Some works have leveraged RQ adjustments for scalability~\cite{wang2013adaptive,grizan2015djay}.
Wang \etal\ presented a system that reduces bandwidth usage by lowering streamed frame resolution, while Grizan \etal\ proposed dJay, which adjusts RQ to reduce GPU resource usage.
While dJay shares similarity with \sys\ in using RQ, \sys\ has more advanced approaches with its user-perceived quality-aware RQ optimization, efficiency-based prioritization, and RQ stabilization mechanisms (more details in Appendix~\ref{sec:vs_djay}).
In our evaluation, we demonstrate the effectiveness of \sys's features for gaming service quality compared to existing approaches including dJay.

\section{Conclusion}
\label{sec:conclusion}

This paper introduced a new insight: the computational effort for real-time content generation in cloud gaming can be wasted because the resulting high-quality output becomes ineffective on the user side due to network-induced lossy compression.
Building on this insight, we developed \sys, a novel system that adaptively optimizes RQ by balancing server-side costs with user-perceived visual quality.
\sys's efficiency-driven approach significantly improves server scalability, allowing it to serve more users while ensuring a stable and satisfactory experience.
Our comprehensive evaluations and a user study confirmed that \sys\ provides a measurably better and more consistent user experience than existing methods, validating its effectiveness beyond just performance metrics.

\section*{Acknowledgment}
\label{sec:ack}
We thank our shepherd, Zili Meng, and the anonymous reviewers for their invaluable feedback and help in improving the paper.
This work has been partially supported by NSF projects CCF-2217070 and CNS-1909769, the Applications Driving Architectures (ADA) Research Center, a JUMP Center co-sponsored by SRC and DARPA, and by funding and equipment gifts from VMware and Intel, and travel support from Dolby Laboratories.

\bibliographystyle{plain}
\bibliography{ref}
\newpage
\appendix
\section{Trade-offs in RQ Control Granularity}
\label{sec:rq_granularity}
\sys\ currently employs coarse-grained RQ adjustments by leveraging predefined game engine presets, as discussed in our background analysis (\S\ref{sec:background}).
While this approach to the rendering pipeline enables the high cost-reduction potential necessary for maximizing per-GPU scalability, it introduces reconfiguration overhead.
As observed in \S\ref{sec:rq_oscillation}, this can lead to transient FPS down-spikes during RQ transitions.
In contrast, fine-grained quality adjustments at later pipeline stages, such as Variable Rate Shading (VRS)~\cite{vaidyanathan2014coarse}, could offer smoother RQ transitions with lower overhead.
However, such localized adjustments are inherently limited in their ability to reduce overall rendering costs compared to the broader pipeline-level optimizations which \sys\ utilizes.
Thus, a trade-off exists between the granularity of RQ control, the achievable resource savings, and the associated transition overhead.
While this work mainly focuses on demonstrating the benefits of user-perceived quality-aware resource optimization through coarse-grained adjustments, identifying which stages of the rendering pipeline offer the efficient balance of adjustment overhead, cost-reduction capacity, and user-perceived quality remains an area for future investigation.


\section{Model Generalization and Practical Deployment Strategy}
\label{sec:cross_validation}

Stimpack optimizes resource efficiency by assigning RQ based on a predictive model of user-side frame quality.
This section discusses the model's generalization capabilities and outlines a deployment strategy tailored for production cloud gaming environments.

\summary{Model Robustness with Cross Validation.}\quad
To evaluate the robustness of our quality prediction model, we performed cross-validation across the collected dataset.
Following the methodology in \S\ref{sec:frameprediction}, we partitioned the data by game scene locations, training the model on specific subsets and testing it on unseen scenes.
This process was iterated across all scene combinations.

\begin{figure}[t]
  \centering
  \vspace{-2.0ex}
  \includegraphics[width=0.8\linewidth]{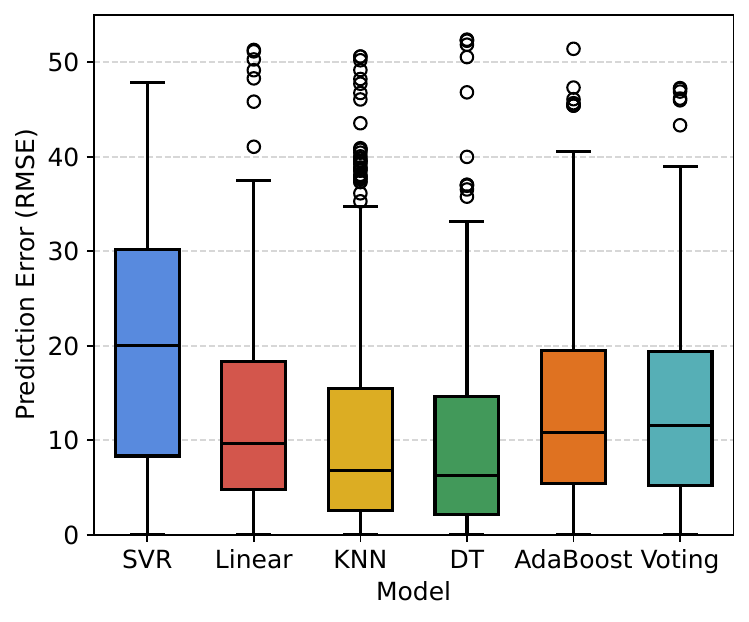}
  \vspace{-2.0ex}
  \caption{The distribution of prediction errors (RMSE) from cross-validation for different prediction models}
  \label{fig:cv_models}
  \vspace{-2.0ex}
\end{figure}

\begin{figure}[t]
  \centering
  \begin{subfigure}[t]{0.23\textwidth}
    \includegraphics[width=\textwidth]{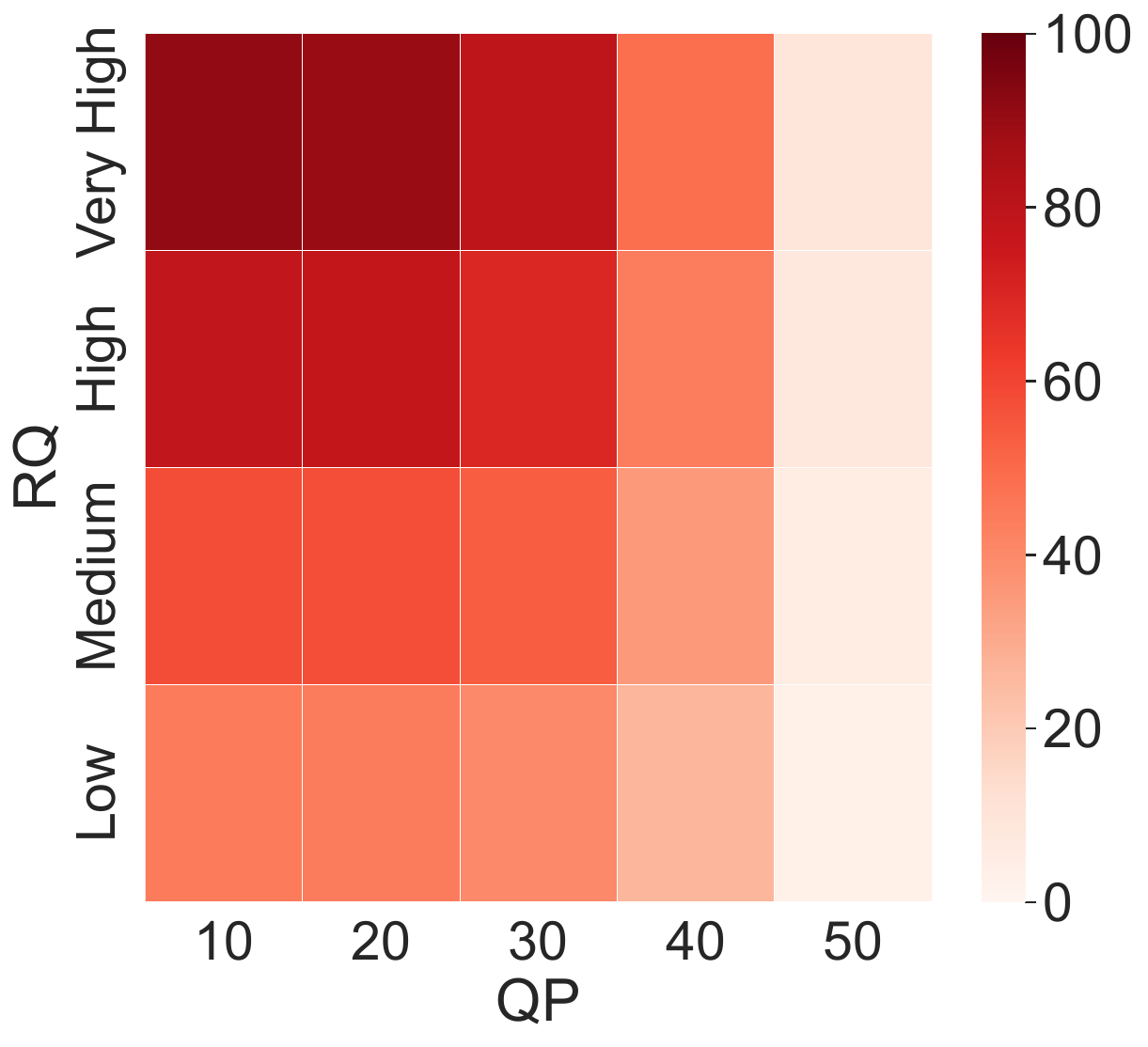}
    \vspace{-3.0ex}
    \caption{Forest}
    \label{fig:heatmap1}
  \end{subfigure}\hspace{0.01\textwidth}
  \begin{subfigure}[t]{0.23\textwidth}
    \includegraphics[width=\textwidth]{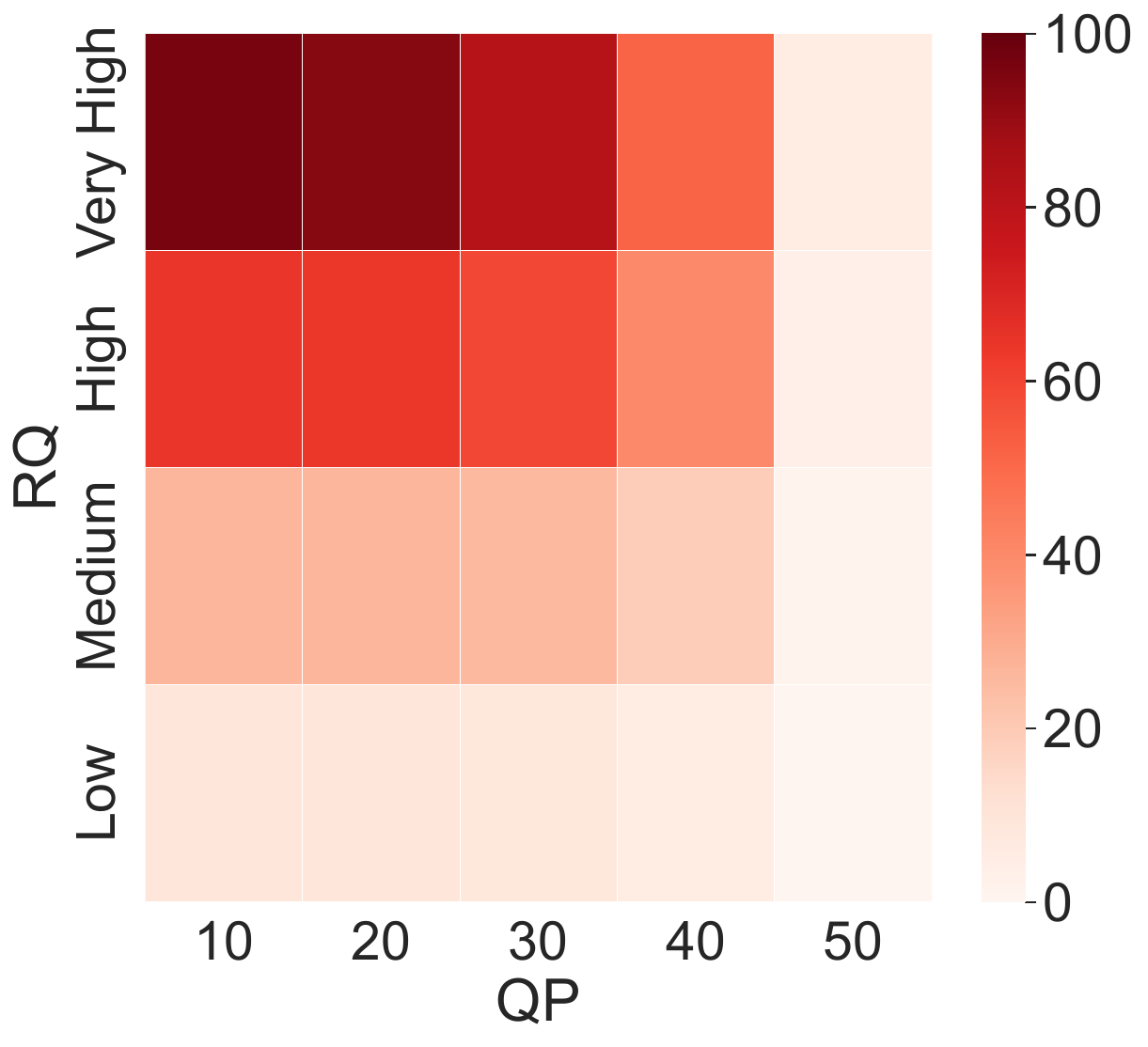}
    \vspace{-3.0ex}
    \caption{Sky Field}
    \label{fig:heatmap2}
  \end{subfigure}
  \vspace{-1.0ex}
  \caption{The VMAF heatmaps with different RQs and QPs in two sample scenes}
  \vspace{-2.0ex}
  \label{fig:heatmap}
\end{figure}

Figure~\ref{fig:cv_models} illustrates the prediction error (RMSE) distribution for various regression models (SVR, Linear, KNN, Decision Tree, AdaBoost, and Voting).
We observe a trade-off: models with higher overall errors (SVR, Linear, AdaBoost, Voting) produce fewer outliers, whereas those with lower median errors (KNN, DT) exhibit more frequent outliers.
This indicates that while KNN and DT perform well on average, they can fail on certain scenes with unique characteristics, \eg, graphics complexities or motion dynamics, that deviate from the training distribution.
This observation is corroborated by the average VMAF heatmaps in Figure~\ref{fig:heatmap}; even under identical RQ and QP settings, the Forest and Sky Field scenes of Figure~\ref{fig:sc} yield different quality changes, highlighting how scene-specific characteristics impact prediction performance.

\summary{Practical Deployment Strategy.}\quad
These findings suggest that a single general model may not suffice for all scenarios, necessitating a trade-off between model generalization and specialization.
To navigate this in production environments, we envision a hybrid deployment strategy.

In the video game industry, a few blockbuster AAA titles command extensive player bases, while a long tail of diverse, less popular games exists.
For these high-impact titles, the significant QoE gains from specialized models can justify the additional costs of per-game training and maintenance.
Conversely, the vast long tail of titles can be served by a general model, ensuring broad system scalability without the need for exhaustive per-title optimization.

Beyond the parameters investigated in this study (RQ and QP), prediction accuracy could be further refined by incorporating content-aware features with advanced prediction methods.
For instance, integrating scene metadata such as motion intensity, texture complexity, or even user-side display characteristics could provide a more granular understanding of perceived quality.
While exploring these high-dimensional feature sets is beyond the scope of this paper, such approaches would enhance the prediction model's robustness across diverse game play scenarios.


\section{Metric Validity: Current Approach and Comprehensive QoE}
\label{sec:metric_validity}
\summary{Rationale for Service Quality Score.}\quad
To the best of our knowledge, a unified and standardized metric that fully captures cloud gaming QoE does not currently exist~\cite{metzger2022introduction}.
Traditionally, gaming performance has been evaluated using disjoint metrics, primarily visual quality (\eg, PSNR, SSIM, VMAF) and smoothness (\eg, FPS, frame time).
However, evaluating these dimensions in isolation fails to capture the holistic user experience.
To address this gap, we designed the service quality score (Eq.~\ref{eq:sqs}) by combining VMAF and a logarithmic FPS score.
The logarithmic scaling of FPS reflects the Weber-Fechner law of the human visual system, where the perceived benefit of higher frame rates diminishes as FPS increases~\cite{reichl2010logarithmic}.
This fused metric facilitates the quantitative comparison of Stimpack against existing baselines by providing a single, representative value for overall service quality.

\summary{The Need for Holistic QoE Metrics.}\quad
We acknowledge that the service quality score is a proxy and does not yet encompass all factors influencing user experience, particularly network performance metrics inherent to cloud gaming, such as jitter, packet loss, and latency.
As VMAF successfully fused multiple metrics into a comprehensive perceptual quality score through extensive validation, similar efforts are required to develop such QoE metrics that incorporate both gaming and network dimensions.
We argue that the current lack of such comprehensive metrics acts as a constraint, limiting the development of more efficient cloud gaming systems and the ability to quantify system design trade-offs.

\summary{Towards Application-Specific Efficiency.}\quad
Stimpack positions itself as a stepping stone toward this goal.
In this work, we demonstrate that incorporating application-level awareness into resource optimization leads to superior user-perceived quality while simultaneously improving scalability.
This highlights the relationship between metric design and system efficiency.
We argue that the development of more reliable, holistic QoE metrics, incorporating visual fidelity, smoothness, and network interactivity, will unlock new levels of efficiency for cloud gaming systems, and we leave the integration of these factors as a future research direction.


\section{Performance under Real-World Dynamics and Stress}
\label{sec:app_dynamics_stress}

\subsection{Adaptability to Dynamic Network Conditions}
To validate \sys's effectiveness in realistic cloud gaming scenarios, we conducted additional experiments using dynamic network conditions.
In our evaluations (\S\ref{sec:evaluation}), even when considering multiple users with different QPs, we assumed their QP values remained constant throughout the session.
While this synthetic setup was beneficial for evaluating \sys's core mechanisms, it does not fully represent real-world cloud gaming environments, where users' network conditions fluctuate over time, directly impacting the compression levels applied to their video streams.

\begin{figure}[t]
  \centering
  \begin{subfigure}[t]{0.5\textwidth}
    \includegraphics[width=\textwidth]{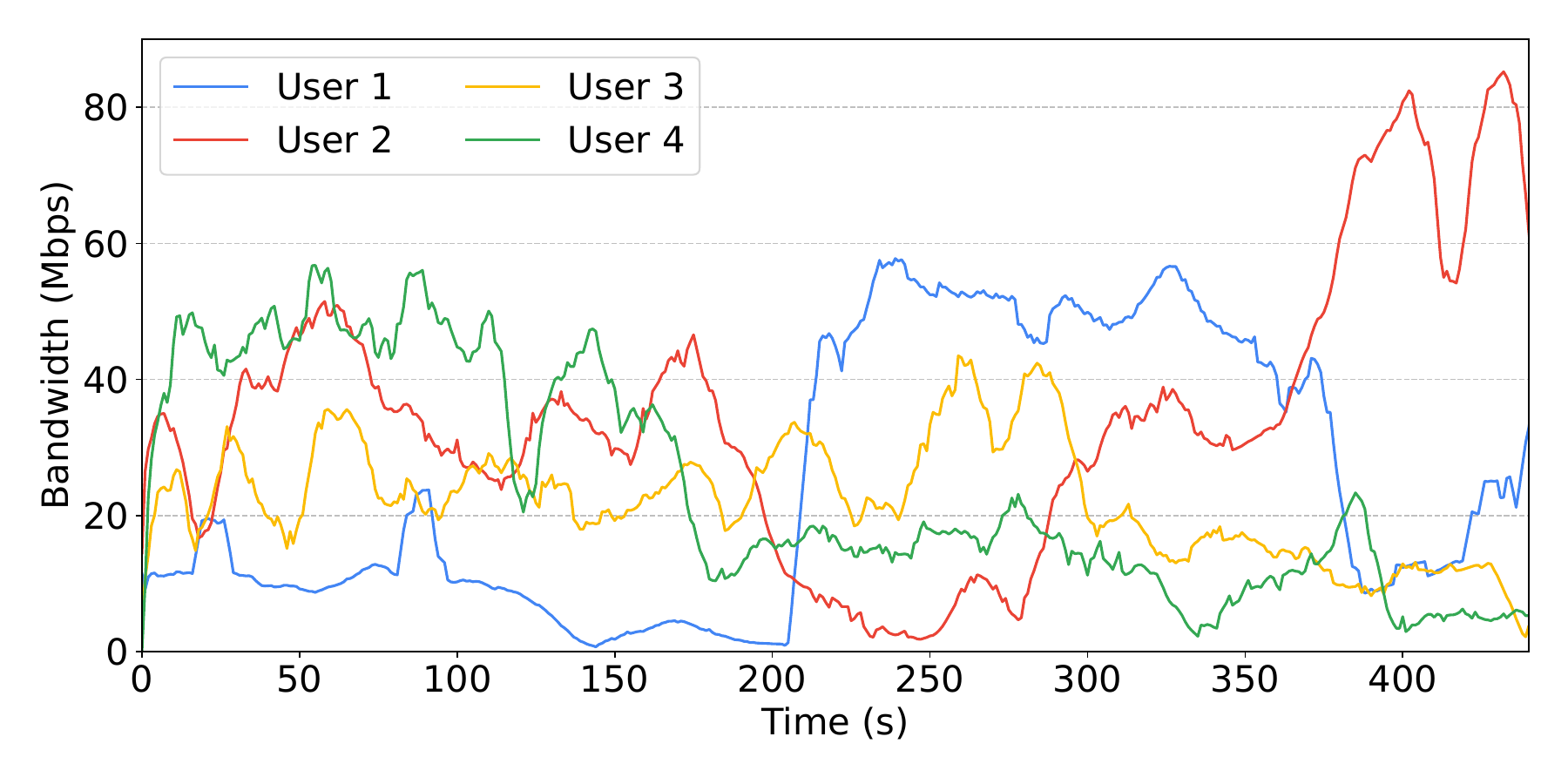}
    \vspace{-5.0ex}
    \caption{The selected 4G/LTE bandwidth traces for 4 users}
    \label{fig:app_mbps}
  \end{subfigure}\hspace{0.01\textwidth}
  \begin{subfigure}[t]{0.5\textwidth}
    \includegraphics[width=\textwidth]{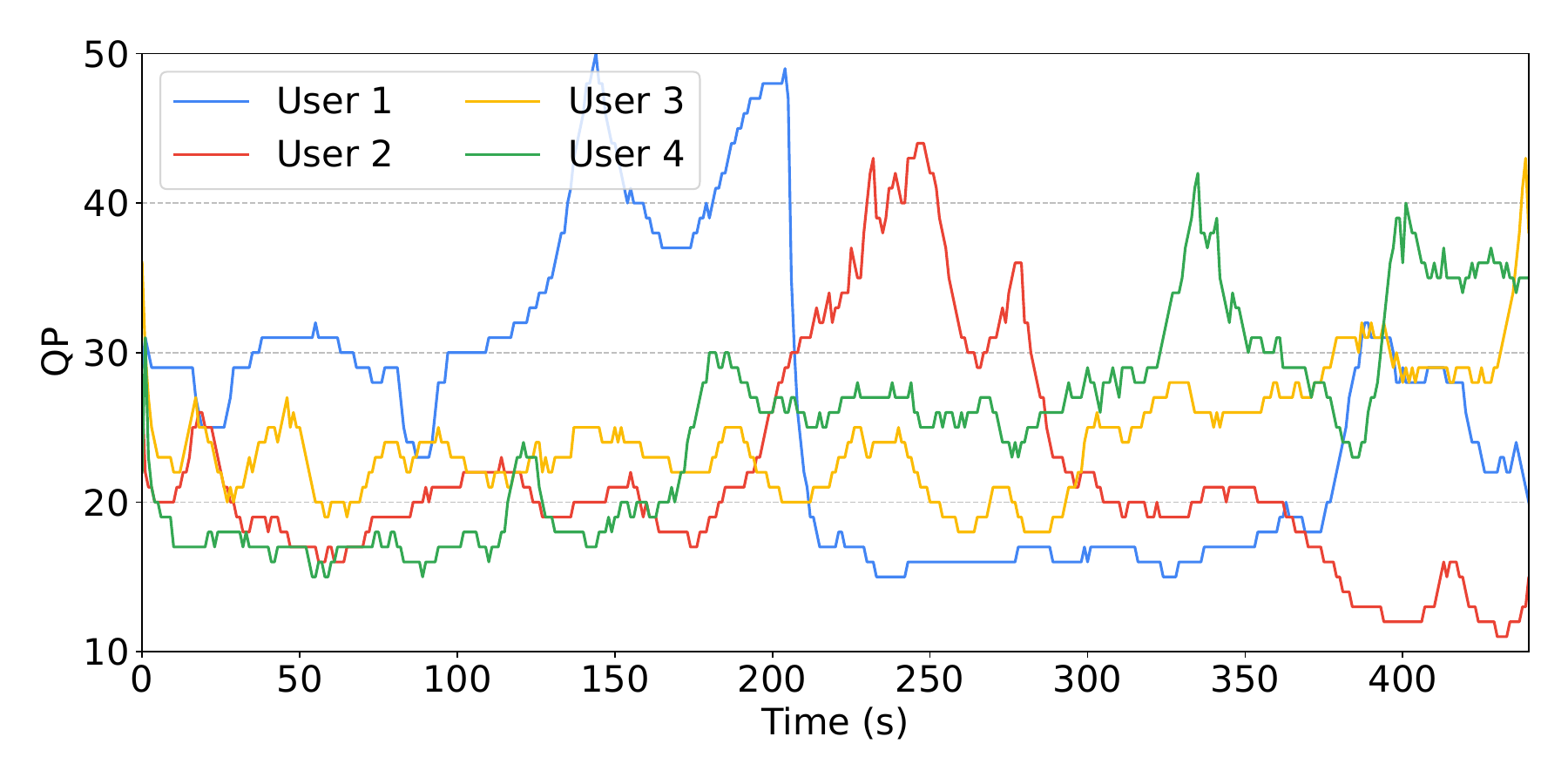}
    \vspace{-5.0ex}
    \caption{The generated QP traces from the bandwidth traces}
    \label{fig:app_qp}
  \end{subfigure}
  \vspace{-1.0ex}
  \caption{The bandwidth and QP traces used in the dynamic network condition experiment}
  \vspace{-2.0ex}
  \label{fig:app_mbps_qp}
\end{figure}

\summary{Experimental Setup.}\quad
To reflect real-world network dynamics, we designed an experiment where each user's QP trace varies based on actual 4G/LTE bandwidth measurements~\cite{vanderHooft2016}.
We selected four distinct bandwidth traces (Figure~\ref{fig:app_mbps}) for four concurrent users playing Village Shooter.
By compressing a gameplay clip with FFmpeg~\cite{ffmpeg} to match these bandwidth logs, we generated corresponding QP traces (Figure~\ref{fig:app_qp}).
As expected, the QP traces exhibit an inverse relationship with bandwidth: lower bandwidth necessitates a higher QP (more lossy compression), while higher bandwidth allows for better frame quality with a lower QP.

\begin{figure}[t]
  \centering
  \includegraphics[width=\linewidth]{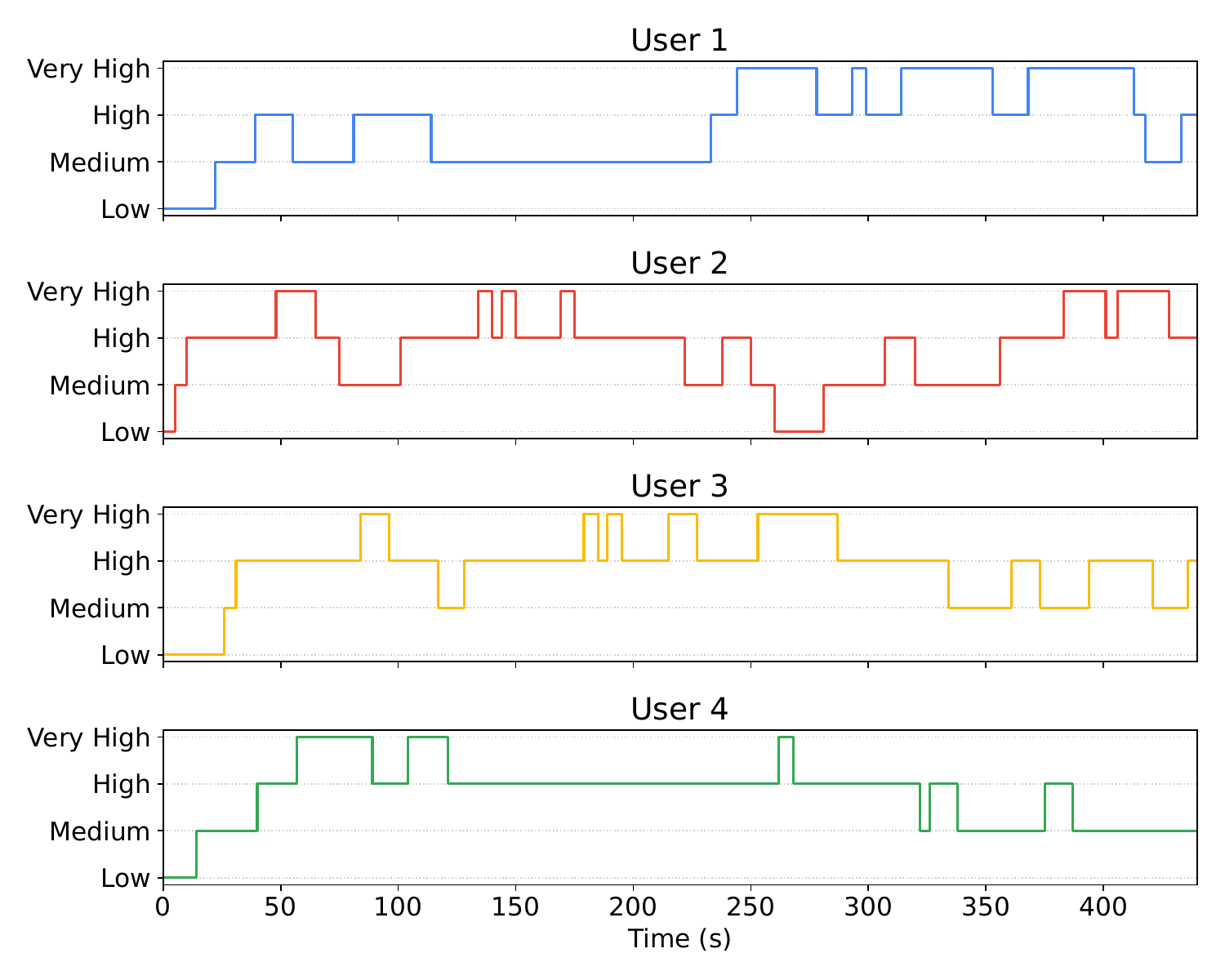}
  \vspace{-5.0ex}
  \caption{The RQ traces for 4 users under dynamic network conditions}
  \label{fig:rq_faceted}
\end{figure}

\begin{figure}[t]
  \centering
  \includegraphics[width=\linewidth]{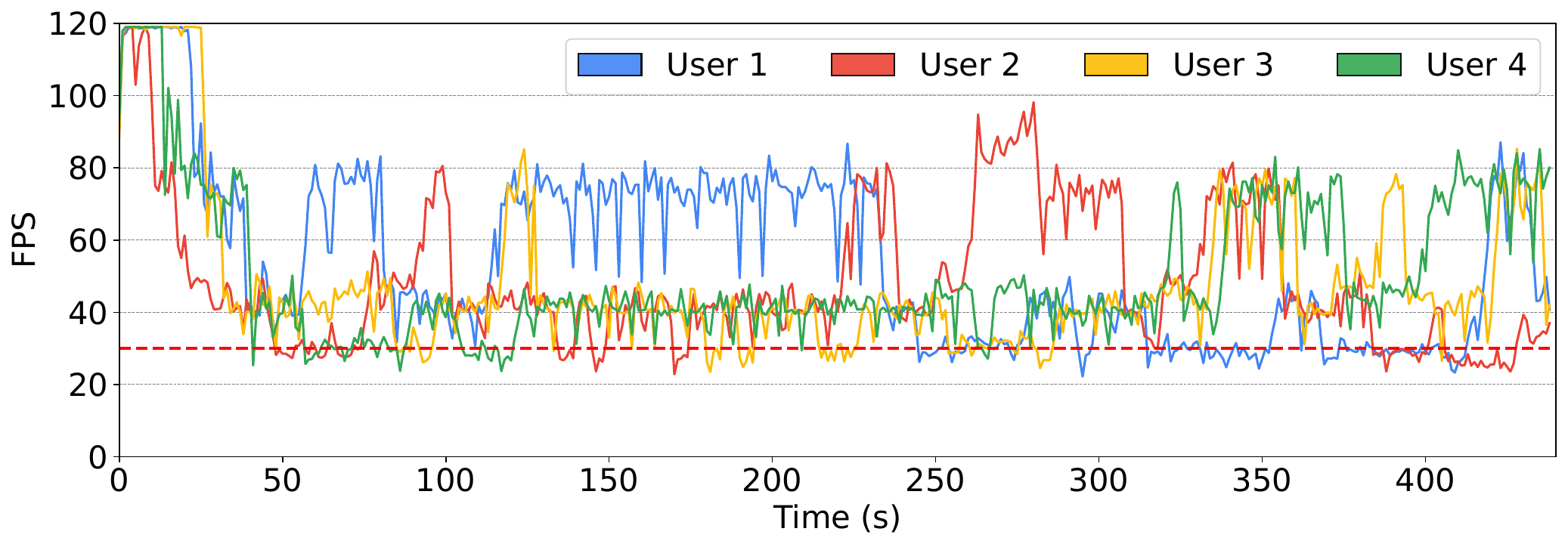}
  \vspace{-4.0ex}
  \caption{The FPS traces for 4 users under dynamic network conditions}
  \label{fig:net_fps_res}
\end{figure}

\summary{Results and Analysis.}\quad
We measured how \sys\ optimizes RQs for each user under these dynamic QP conditions.
Figure~\ref{fig:rq_faceted} shows the RQ traces for the four users.
A key observation is that each user's RQ level closely mirrors their bandwidth trace in Figure~\ref{fig:app_mbps}, demonstrating that \sys\ effectively adapts RQs in response to changing network conditions with prioritization.
Specifically, User 1 is maintained at a \texttt{Medium} RQ during an initial period of poor network condition (\tlide200 s), while the others are prioritized with higher RQs due to their superior bandwidth.
Around \tlide 200 s, as User 1's network condition improves and User 2's degrades, \sys\ dynamically adjusts their RQs by increasing User 1's RQ to \texttt{Very High} and lowering User 2's to \texttt{Low}.
Later, around \tlide 400 s, as User 2's condition recovers and the others worsen, \sys\ again reallocates resources, elevating User 2's RQ to \texttt{Very High} while reducing the others.
These adjustments demonstrate that \sys\ adaptively reallocates resources based on real-time network conditions, effectively identifying which users can benefit most from higher rendering quality.
Furthermore, Figure~\ref{fig:net_fps_res} confirms that \sys\ maintains a playable FPS for all users throughout these transitions.

\subsection{Limitations and Cluster-Level Orchestration}
While \sys\ improves per-GPU scalability, our evaluations revealed its operational boundaries under extreme stress.
When a single server is subjected to excessive load, \sys\ must prioritize basic playability by aggressively reducing RQs across all sessions.
As shown with Scenario 2 of Table~\ref{tab:scal}, the opportunity for \sys's to adjust RQs becomes limited in such saturated states, as there is insufficient GPU budget to promote a user's RQ without compromising others' playability.
These observations indicate that under extreme overload, localized optimization alone cannot overcome the resource constraints.

Such results imply that the performance gains from \sys\ are most effective when the system has sufficient operational headroom to make meaningful trade-offs, necessitating its integration into a more comprehensive system architecture.
In a practical production environment, \sys\ is intended to function as a optimization layer that manages per-GPU efficiency for individual servers.
To achieve end-to-end reliability, it should be paired with cluster-level orchestration that operates at a higher hierarchy.
This management layer would handle workload distribution and admission control mechanisms to prevent individual servers from reaching the saturation points where \sys's optimization capacity becomes limited.


\section{Qualitative Comparison: Paradigm Shift from Prior Work}
\label{sec:vs_djay}
While prior works like dJay~\cite{grizan2015djay} have successfully demonstrated the potential of adjusting RQ to conserve GPU resources, Stimpack represents a fundamental shift from \textit{server-centric resource management} to \textit{end-to-end experience optimization}.
We summarize the qualitative differences in Table~\ref{tab:comparison}.

\summary{Compression-Awareness vs. Network-Blindness.}\quad
The most critical distinction lies in how the system perceives quality.
dJay operates in a network-blind manner; it assumes that higher rendering quality on the server is translated to better visual quality on the user side, provided the target FPS is met.
However, as our results show (Figure~\ref{fig:motiv} and~\ref{fig:motiv_rq_cost}), this assumption is not always valid in cloud gaming scenarios where the rendered frames are not transmitted losslessly.
High-fidelity frames rendered by dJay may be severely degraded by the video encoder when the bitrate is insufficient, resulting in wasted GPU resources that do not contribute to the user's effective experience.
In contrast, Stimpack is compression-aware.
By explicitly modeling the interaction between RQs and QPs, Stimpack identifies the point of diminishing returns.
It recognizes when a user's network condition cannot sustain high detail and lower rendering effort, redirecting those resources to users who can actually perceive the improvement.

\summary{Efficiency-Based Prioritization.}\quad
Consequently, the resource allocation strategy using RQ differs fundamentally.
While dJay typically applies uniform adjustments across users, Stimpack presents an efficiency-based prioritization scheme.
It treats GPU resources as a shared budget and allocates them based on the how effectively resource usage is translated into user-perceived quality improvements.
This efficiency-driven approach enables Stimpack to maintain the higher overall service quality, while still accommodating multiple users under same resource footprint.

\begin{table}[h]
  \centering
  \caption{The comparison summary between dJay and Stimpack}
  \vspace{-2.0ex}
  \label{tab:comparison}
  \resizebox{\columnwidth}{!}{%
    \begin{tabular}{@{}l|l|l@{}}
      \toprule
      \textbf{Feature / Aspect} & \textbf{dJay} & \textbf{Stimpack} \\ \midrule
      \textbf{Optimization Goal} & \begin{tabular}[c]{@{}l@{}}Server-centric resource saving \\ for playable FPS\end{tabular} & \begin{tabular}[c]{@{}l@{}} Maintaining user-perceived \\ quality with playable FPS \end{tabular} \\ \midrule
        \textbf{Network Awareness} & \begin{tabular}[c]{@{}l@{}}Network-blind\\ (Ignores compression loss)\end{tabular} & \begin{tabular}[c]{@{}l@{}}Compression-aware\\ (Considers user-side quality)\end{tabular} \\ \midrule
          \textbf{Resource Allocation} & \begin{tabular}[c]{@{}l@{}}FPS-driven adjustment \\ (Uniform across users) \end{tabular} & \begin{tabular}[c]{@{}l@{}}Efficiency-based prioritization\end{tabular} \\ \midrule
            \textbf{Stability Mechanism} & \begin{tabular}[c]{@{}l@{}}Immediate adjustment\\ (Prone to oscillation)\end{tabular} & \begin{tabular}[c]{@{}l@{}}Backoff to mitigate oscillation \end{tabular} \\ \bottomrule
    \end{tabular}%
    }
  \vspace{-1.0ex}
\end{table}

\summary{Stability and Practicality.}\quad
Finally, Stimpack highlights the importance of stability in RQ adjustments for user experience and incorporates a backoff mechanism to diminish oscillations.
While dJay adjusts RQ every round based on immediate FPS feedback, Stimpack's stabilization mechanism prevents rapid FPS fluctuations that can be more detrimental to user experience than static low quality, which is also supported by our user study (\S\ref{sec:userstudy}).
This holistic approach makes Stimpack not just a resource optimizer, but a comprehensive QoE management system for cloud gaming.

\end{document}